\documentclass[pdftex,twocolumn,epjc3]{svjour3}          

\RequirePackage[T1]{fontenc}

\smartqed  

\RequirePackage{graphicx}
\RequirePackage{mathptmx}      
\RequirePackage{flushend}
\RequirePackage[numbers,sort&compress]{natbib}
\RequirePackage[colorlinks,citecolor=blue,urlcolor=blue,linkcolor=blue]{hyperref}
\RequirePackage{amsmath}
\RequirePackage{amsfonts}
\RequirePackage{bbm}
\RequirePackage{caption} 
\RequirePackage{color}
\RequirePackage{enumitem} 
\RequirePackage{float}
\RequirePackage{graphics}
\RequirePackage{physics} 
\RequirePackage{subcaption} 
\RequirePackage{textgreek}
\RequirePackage{breqn}
\usepackage{lineno}

\journalname{Eur. Phys. J. C}

\newcommand{\addOneFig}[4]
{
\begin{figure}[htp]
  \begin{center}
    \includegraphics[width=#4\textwidth]{#1}
    \caption{#2}
    \label{#3}
  \end{center}
\end{figure}
}

\newcommand{\addTwoFigs}[9]
{

\begin{figure}[htp]

\begin{subfigure}[t]{#9\textwidth}
    \includegraphics[width=\textwidth]{#1}
    \caption{#2}
    \label{#3}
\end{subfigure}\hspace{\fill} 
\begin{subfigure}[t]{#9\textwidth}
    \includegraphics[width=\linewidth]{#4}
    \caption{#5}
    \label{#6}
\end{subfigure}

\caption{#7}
\label{#8}
\end{figure}
}

\newcommand{\addTwoFigsH}[9]
{

\begin{figure*}[htp]

\begin{subfigure}[t]{#9\textwidth}
    \includegraphics[width=\textwidth]{#1}
    \caption{#2}
    \label{#3}
\end{subfigure}\hspace{\fill} 
\begin{subfigure}[t]{#9\textwidth}
    \includegraphics[width=\linewidth]{#4}
    \caption{#5}
    \label{#6}
\end{subfigure}

\caption{#7}
\label{#8}
\end{figure*}
}

\newcommand{\addThreeFigsH}[9]
{
  \def\pathImgA{#1}
  \def\captImgA{#2}
  \def\labelImgA{#3}
  
  \def\pathImgB{#4}
  \def\captImgB{#5}
  \def\labelImgB{#6}
  
  \def\pathImgC{#7}
  \def\captImgC{#8}
  \def\labelImgC{#9}
  \addThreeFigsTemplateH
}

\newcommand{\addThreeFigsTemplateH}[3]
{
\begin{figure*}[htp]
\centering

\subfloat[\captImgA]{%
  \includegraphics[height=#3\textheight]{\pathImgA}%
  \label{\labelImgA}%
}\qquad
\subfloat[\captImgB]{%
  \includegraphics[height=#3\textheight]{\pathImgB}%
  \label{\labelImgB}%
}\qquad
\subfloat[\captImgC]{%
  \includegraphics[height=#3\textheight]{\pathImgC}%
  \label{\labelImgC}%
}

\caption{#1}
\label{#2}
\end{figure*}
}
\newcommand{\addFourFigs}[9]
{
  \def\pathImgA{#1}
  \def\captImgA{#2}
  \def\labelImgA{#3}
  
  \def\pathImgB{#4}
  \def\captImgB{#5}
  \def\labelImgB{#6}
  
  \def\pathImgC{#7}
  \def\captImgC{#8}
  \def\labelImgC{#9}
  \addFourFigsTemplate
}

\newcommand{\addFourFigsTemplate}[6]
{
\begin{figure*}[htp]

\begin{subfigure}[t]{#6\textwidth}
    \includegraphics[width=\textwidth]{\pathImgA}
    \caption{\captImgA}
    \label{\labelImgA}
\end{subfigure}\hspace{\fill} 
\begin{subfigure}[t]{#6\textwidth}
    \includegraphics[width=\linewidth]{\pathImgB}
    \caption{\captImgB}
    \label{\labelImgB}
\end{subfigure}

\bigskip 
\begin{subfigure}[t]{#6\textwidth}
    \includegraphics[width=\linewidth]{\pathImgC}
    \caption{\captImgC}
    \label{\labelImgC}
\end{subfigure}\hspace{\fill} 
\begin{subfigure}[t]{#6\textwidth}
    \includegraphics[width=\linewidth]{#1}
    \caption{#2}
    \label{#3}
\end{subfigure}

\caption{#4}
\label{#5}
\end{figure*}
}

\renewcommand{\deg}[0]{\circ}
\newcommand{\kH}{\ket{\text{H}}}
\newcommand{\kV}{\ket{\text{V}}}

\newcommand{\kHV}{\ket{\text{HV}}}
\newcommand{\kVH}{\ket{\text{VH}}}

\newcommand{\bHV}{\bra{\text{HV}}}

\newcommand{\kbHH}{\ketbra{\text{H}}{\text{H}}}
\newcommand{\kbVV}{\ketbra{\text{V}}{\text{V}}}
\newcommand{\kbHV}{\ketbra{\text{H}}{\text{V}}}

\newcommand{\re}{\frac{r_{0}^{2}}{2}}
\newcommand{\tTheta}{\tilde{\Theta}}
\newcommand{\eps}{\boldsymbol \varepsilon}
\newcommand{\Kr}{\mathcal{K}} 
\newcommand{\id}{\mathbbm{1}}
\newcommand{\Pauli}[1]{\hat{\sigma}_{#1}}
\newcommand{\uv}[1]{\hat{\bf{#1}}}
\newcommand{\fF}[1]{\mathcal{F}(#1)}
\newcommand{\fV}[1]{\mathcal{V}(#1)}
\newcommand{\sF}{\mathcal{F}}
\newcommand{\sV}{\mathcal{V}}
\newcommand{\DCS}[1]{\frac{d\sigma_{#1}}{d\Omega}}

\newcommand{\Vis}{\mathcal{V}}

\newcommand{\TR}[1]{\text{Tr}\left[#1\right]}
\newcommand{\IM}[1]{\text{Im}(#1)}

\newcommand{\DeltaPhi}{\Delta\hat{\Phi}}
\newcommand{\Mean}[1]{\langle #1 \rangle}

\newcommand{\OneLineEq}[2]{
\begin{equation}\label{#1}
  #2
\end{equation}
}

\newcommand{\ManyLinesEq}[2]{
\begin{dmath}\label{#1}
  #2
\end{dmath}
}

\newcommand{\CManyLinesEq}[2]{
\begin{multline}\label{#1}
  #2
\end{multline}
}

\newcommand{\refFig}[1]{Fig.~\ref{#1}}

\newcommand{\refEq}[1]{Eq.~\ref{#1}}
\newcommand{\refSec}[1]{Sect. \ref{#1}}
\newcommand{\refPub}[1]{~\cite{#1}}

\begin{document}

\title{Probing arbitrary polarized photon pairs undergoing double Compton scatterings by a dedicated MC simulator validated with experimental data}

\author{
Mateusz Bała\thanksref{addr_DCS_NCBJ}
\and
Wojciech Krzemień\thanksref{addr_HEPD_NCBJ} \thanks{\emph{corresponding author:} wojciech.krzemien@ncbj.gov.pl} 
\and
Beatrix C. Hiesmayr\thanksref{addr_UVFP}
\and
Jakub Baran\thanksref{addr_FAIS}
\and
Kamil Dulski\thanksref{addr_FAIS}
\and
Konrad Klimaszewski\thanksref{addr_DCS_NCBJ}
\and
Lech Raczyński\thanksref{addr_DCS_NCBJ}
\and
Roman Y. Shopa\thanksref{addr_DCS_NCBJ}
\and
Wojciech Wiślicki\thanksref{addr_DCS_NCBJ}
}

\institute{Department of Complex Systems, National Centre for Nuclear Research, Andrzeja Soltana 7, Otwock, Swierk, PL-05-400, Poland\label{addr_DCS_NCBJ}
\and
High Energy Physics Division, National Centre for Nuclear Research, Andrzeja Soltana 7, Otwock, Swierk, PL-05-400, Poland\label{addr_HEPD_NCBJ}
\and
University of Vienna, Faculty of Physics, Währingerstrasse 17, Vienna, 1090, Austria\label{addr_UVFP}
\and 
Marian Smoluchowski Institute of Physics, Jagiellonian,University, Kraków, Poland\label{addr_FAIS}
}


\maketitle

\begin{abstract}

Quantum correlations in the polarization degrees of freedom of the two-photon system have been extensively studied and form our current understanding of the quantum nature of our world.
Most of the studies are concentrated on the low-energy (optical) photon pairs, for which efficient polarization measurement devices exist. However, for high-energetic (MeV) pairs of photons, e.g. produced in the decay of positronium atoms, no polarizers are available. 
Partial information about the polarization degree of freedom can be extracted by exploiting the measurements of photon pairs that undergo double Compton scattering.  We present a Geant4-based Monte Carlo Vienna-Warsaw model capable of simulating any initial polarization state of bipartite photons.
This puts us in a position to derive the behavior of the experimental observable, the angular difference $\Delta\hat\Phi$ formed by the two scattering planes.
We validate our Vienna-Warsaw simulator with the high-statistics experimental sample --based on a total of $3 \times 10^5 $ event candidates --- of two-photon pairs measured with the J-PET Big Barrel detector. We deduce the value of the squared visibility (interference contrast) encoding the polarization in the angle difference of the two scattering planes, $\Delta\hat\Phi$. The simulated spectra are in good agreement with the experimental correlation spectra and behave as predicted by theory. 
\end{abstract}


\section{Introduction}
\label{sec:introduction}

The physics of photons is very rich, and finding a consistent formalism to describe all phenomena in a unified way is ongoing research. For instance,  since its first proposal by Humblet in 1943\refPub{humblet_sur_1943} the problem of introducing a physically unambiguous separation of the total angular momentum of
photons into a spin part and an orbital part is a controversial and debated subject (e.g.\refPub{singlephoton1,singlephoton2,singlephoton3,singlephoton4,singlephoton5})  with a recently proposed unified treatment via Positive Operator-Valued Measures\refPub{HiesmayrTAM}.  Photon polarisation, a derivative of the spin component, can be considered as an intrinsic photon characteristic connected to its relativistic nature\refPub{HiesmayrTAM}, which is one of the underlying assumptions of this paper.

Polarisation measurements are widely used to show the very working of quantum mechanics and, in particular, quantum correlations of many-photon systems are investigated, providing a way to encode information for quantum information theoretic tasks such as e.g. quantum computing\refPub{madsen_quantum_2022}, quantum machine learning\refPub{MLHiesmayr, biamonte_quantum_2017,wan_quantum_2017} or quantum cryptography\refPub{BB84,E91,SecretSharing,ExpQC,LongDistance}. Most of the studies so far have been concentrated on optical photon systems, in which the energies of the photons are in the range of a few eV. 
However, recent investigations based on the estimation of the polarization for the MeV photon pairs originating from the decay of the positronium atoms or the direct electron-positron annihilation have regained attention. Particularly,  the interest is enhanced by the proposal to exploit it in medical applications, i.e. in the context of positron emission tomography (PET) \refPub{mcnamara_towards_2014, toghyani_polarisation-based_2016,quantum2, bass_colloquium_nodate,wattsQuantumEntangledPET2024}. The main idea is to use the correlation observable - the relative angle of the two-photon scattering planes $\Delta\hat\Phi$ -  to distinguish the genuine, undisturbed photon pairs from the noisy events formed by the uncorrelated (random) photons or photons scattered in the patient's body. The reduction of this background could improve the final PET image contrast\refPub{kozuljevic_study_2021}.  
Indeed, recently several pilot measurements were performed\refPub{makek_single-layer_2020, watts_photon_2021, abdurashitov_setup_2022}. Interestingly, the recent results dedicated to the analysis of the scattered photons \refPub{ ivashkinTestingEntanglementAnnihilation2023,parashariClosingDoorPuzzle2024,bordesFirstDetailedStudy2024} seem to contradict the previous conclusions\refPub{watts_photon_2021}. 
The applicability of the proposed methodology in practical scanners remains, therefore, an open question, nevertheless, further investigations are ongoing~\cite{kimBackgroundReductionPET2023,shoopPETSimulationStudy2024, dasExploringPETImaging2024,kozuljevicInvestigationSpatialResolution2024,wattsQuantumEntangledPET2024,moulinBenefitsQuantumEntanglement2024,moskalNonmaximalEntanglementPhotons2025}.

The polarization of the photon originating from the po\-sitro\-nium decay cannot be determined directly because no perfect polarizer exists.   However,
one can estimate the polarization using the properties of the Compton scattering distribution, which is sensitive to the polarisation degrees of freedom of the photons. A short summary of this method is presented in \refSec{sec:quantum-model} and more details can be found in Ref.\refPub{quantum2}.

In this work,  we present studies of photon correlations probed by the double Compton scattering events based on the theory published in Ref.~\refPub{quantum1}. The investigations are based on the Monte Carlo (MC) simulations utilising the developed `\textit{Vienna-Warsaw model}'(VW) by implementing it
into the Geant4 simulation software. Namely, our VW simulation model allows computing the expected correlation of any initial bipartite state being either separable or entangled, mixed or pure.  
MC-based predictions are validated with the experimental spectra obtained based on the analysis of the high-statistics experimental sample measured by the Jagiellonian Positron Emission Tomograph (J-PET) in 2020~\cite{niedzwiecki_j-pet_2017, dulskiJPETDetectorTool2021a}\footnote{
The experimental input dataset used in this article is overlapping with the one used in~\cite{moskalNonmaximalEntanglementPhotons2025}. The first version of the current article was finalised on 16.08.2023. The submission was postponed at the request of the J-PET group leader P. Moskal, who wanted to prioritize~\cite{moskalNonmaximalEntanglementPhotons2025}.}. 

The rest of the article is structured as follows: the theoretical basis for the VW model is described in~\refSec{sec:quantum-model}. 
We discuss the Compton scattering process in the context of the polarisation measurement in~\refSec{sec:experimental-determination-photon-polarization}.
The details of the VW model are given in \refSec{sec:simulation-model}.
Here we also provide a comparison with the available MC models. 
~\refSec{section:experimental-data} is dedicated to the description of the data analysis scheme and the validation of the MC simulations with the experimental data. 
Finally, the conclusions and further prospects are discussed in~\refSec{sec:summary}.

\section{Prediction of the quantum theory for multi-photonic Compton scattering events}
\label{sec:quantum-model}

We start this section by rewriting the Klein-Nishina formula for any initial polarised photon undergoing Compton scattering in plastic scintillators. A reformulation in Kraus operators allows us to compute the theoretical predictions for any multiphotonic initial state, either separable or entangled, either mixed or pure\refPub{quantum1}. This approach is the basis for the implementation of the MC generator incorporated into the Geant4 software.

\subsection{Klein-Nishina formula for single photons detected by plastic scintillators}
\label{subsec:klein-nishima-single-photon}

For single photon scattering (see \refFig{fig:single-compton}) in a medium Otto Klein and Yoshio Nishina published in 1929 the following formula\refPub{klein_scattering_1928} 

\ManyLinesEq{eq:Klein-Nishina}
{
\DCS{fi} = \re K^2(k,\tTheta)\left(K^{-1}(k,\tTheta)+K(k,\tTheta)-2+4\lvert {\eps_{f}}^{*}\cdot\eps_{i} \rvert^{2}\right)
}
where $i/f$ refer to the initial/final polarised states described by $\varepsilon$, $r_0$ is the classical electron radius, and $k$ is the energy of the incoming photon expressed in $\frac{\text{keV}}{m_{e} c^2}$ (see \refFig{fig:single-compton}), and
\OneLineEq{eq:KN-energy-transfer}{K(k,\tTheta) = \frac{k'(k,\tTheta)}{k}}
where the energy of the outgoing photon scattered by an angle $\tTheta$, the so-called Compton angle, is given by
\OneLineEq{eq:KN-outgoing-energy}{k'(k,\tTheta) =\frac{1}{1-\cos\tTheta+\frac{1}{k}}\;.}

\addOneFig{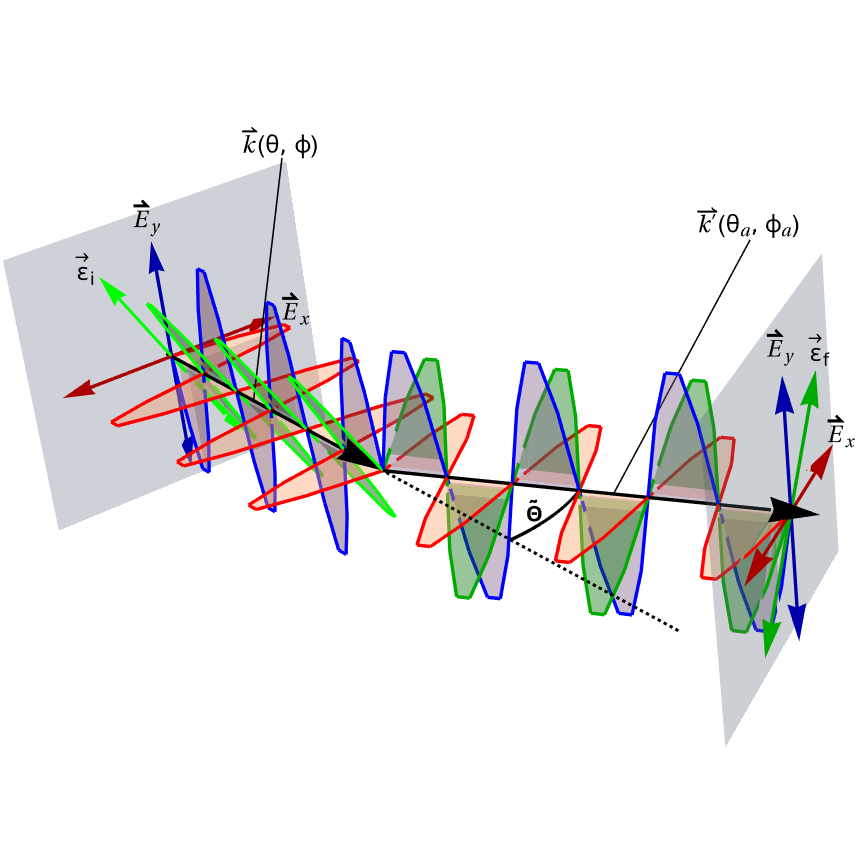}
{The scheme of the Compton scattering of a photon on a free electron. The $\vec{k}$ and $\vec{k'}$ are the wave vectors of the photon before and after scattering, respectively. The initial and final photon polarizations are marked as $\vec{\epsilon}_{i}$ and $\vec{\epsilon}_{f}$, respectively. }
{fig:single-compton}{0.5}

In the case of materials where no preferred orientation of electron spins is defined (e.g., amorphous materials, non-magnetic metals, disordered semiconductors — which would otherwise impose constraints on the photon polarization), it can be assumed that the Compton interaction is not dependent of the outgoing photon polarisation. Consequently, one must sum over all possible final polarization states, i.e. $\DCS{i}=\sum_{f=1}^2 \DCS{fi}$. The material used for detection in the JPET scanner satisfies these conditions — it does not exhibit a preferred spin orientation.

Hence, the Compton scattering events in plastic scintillators can be described by the differential scattering cross section depending on the initial polarised state $\rho$ derived in \refPub{quantum1}:
\OneLineEq{eq:cross-section-one-photon-kraus}
{
 \frac{d}{d\Omega} \sigma_\rho (\theta,\phi,\theta_a,\phi_a,k)
 = \re\; K^2(k,\tTheta) \;\sum_{i=1}^{2} \TR{\Kr_i\rho\Kr_i^\dagger}
}
where the angles $\theta,\phi$ and $\theta_a,\phi_a$ are the angles describing the direction of motion (before and after the Compton scattering) with respect to one chosen coordinate system.
The operators $\Kr_i$ are pseudo-Kraus operators which have to be defined with respect to the chosen basis of the state $\rho$. This is the relation of the internal degrees of freedom (polarisation) with the outer degrees of freedom (position). If we choose the linear polarisation basis $\{H,V\}$, then its directions correspond to the oscillation direction of the electromagnetic field vector in the classical limit. Let us emphasize that in quantum mechanics the electromagnetic field vector is not needed for the definition of polarisation, rather it has to be postulated that every photon has the property `polarisation' as the property 'energy' as well as ``the propagation direction'', however, note that those three quantum numbers are enough to fully describe the photon as a quantum mechanical object\refPub{singlephoton1,singlephoton2,singlephoton3,singlephoton4,singlephoton5,HiesmayrTAM}. 

Any initial state $\rho$ can be written by an orthonormal basis ($\langle H|V\rangle=0,\langle H|H\rangle=\langle V|V\rangle=1$)
\ManyLinesEq{eq:init-state-rho}{\rho= \rho_{HH}\kbHH+\rho_{HV}\kbHV+\rho_{VH}\ket{V}\bra{H} +\rho_{VV}\kbVV}
where $\rho_{HH},\rho_{VV}$ are positive numbers that add up to $\rho_{HH}+\rho_{VV}=1$ and $\rho_{HV}=\rho_{VH}^*$ are complex numbers and generally the positivity $\rho\geq 0$ has to hold. Given this basis choice, the Kraus operators decomposed into Pauli matrices $\Pauli{i}$ are defined by
\OneLineEq{eq:kraus-operator-1}
{
\Kr_1 = \sqrt{\frac{k}{k'(k,\tTheta)}+\frac{k'(k,\tTheta)}{k}-2}\quad\id_2
}
\ManyLinesEq{eq:kraus-operator-2}
{\Kr_2 = \frac{(\cos\theta_a\cos\theta+1)\cos(\phi-\phi_a)+\sin\theta_a\sin\theta}{2}\;\id_2 + \frac{(\cos\theta_a\cos\theta-1)\cos(\phi-\phi_a)+\sin\theta_a\sin\theta}{2}\;\Pauli{3}
+ \frac{i \sin(\phi-\phi_a) (\cos\theta-\cos\theta_a)}{2}\;\Pauli{1} + \frac{\sin(\phi-\phi_a) (\cos\theta+\cos\theta_a)}{2}\;\Pauli{2}\;.}
 The direction of motion and a chosen coordinate system are connected via the Compton scattering angle
\ManyLinesEq{eq:compton-scattering-angle}{\cos\tTheta =\uv{k}\cdot\uv{k_a} = \cos(\theta_a-\theta) + \left(\cos(\phi-\phi_a)-1\right)\sin\theta_a\sin\theta\;.}


Let us rewrite the result for an arbitrary polarised state $\rho$ decomposed into the linear polarisation eigenstates $\{H,V\}$ (for details consult \refPub{quantum1})
\CManyLinesEq{eq:cross-section-probability-single-photon}
{
\frac{d}{d\Omega}\sigma_\rho(\tTheta,\Phi, k)= \re\fF{\tTheta,k}\biggl\{1 -\\ \fV{\tTheta,k}\cdot\biggl((\rho_{HH}-\rho_{VV})\cos(2\Phi) + \IM{\rho_{HV}}\sin(2\Phi)\biggr)\biggr\}
}
with an envelope function (see \refFig{fig:vis_env:env})
\OneLineEq{eq:envelope}
{
\fF{\tTheta,k} = K^2(k,\tTheta)\biggl( K^{-1}(k,\tTheta) + K(k,\tTheta) -\sin^{2}\tTheta\biggr)
}
and a visibility/interference contrast (see \refFig{fig:vis_env:vis})
\OneLineEq{eq:visibility}{
\fV{\tTheta,k} = \frac{\sin^{2}\tTheta}{K^{-1}(k,\tTheta)+K(k,\tTheta)-\sin^{2}\tTheta}\;,
}
where $\Phi = \phi-\phi_a$. 
Consequently, the polarisation vectors within the scattering plane (defined by the vectors $\uv{k}\cdot\uv{k'}$)  have to change by $\cos\tTheta$ and the vectors orthogonal to this plane are left unchanged since polarisation is defined orthogonal to the propagation. 

\addTwoFigsH
{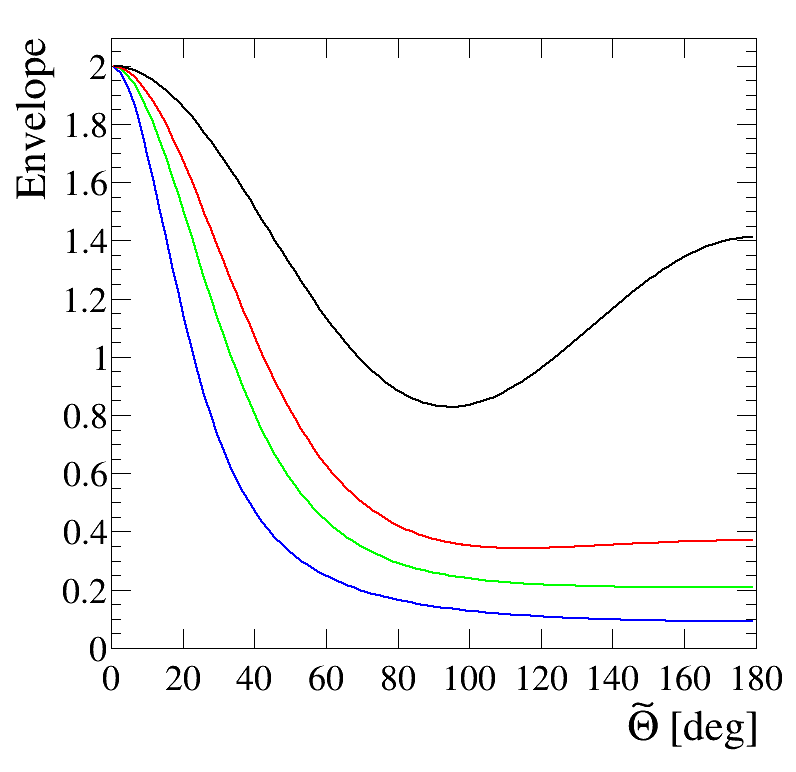}{}{fig:vis_env:env}
{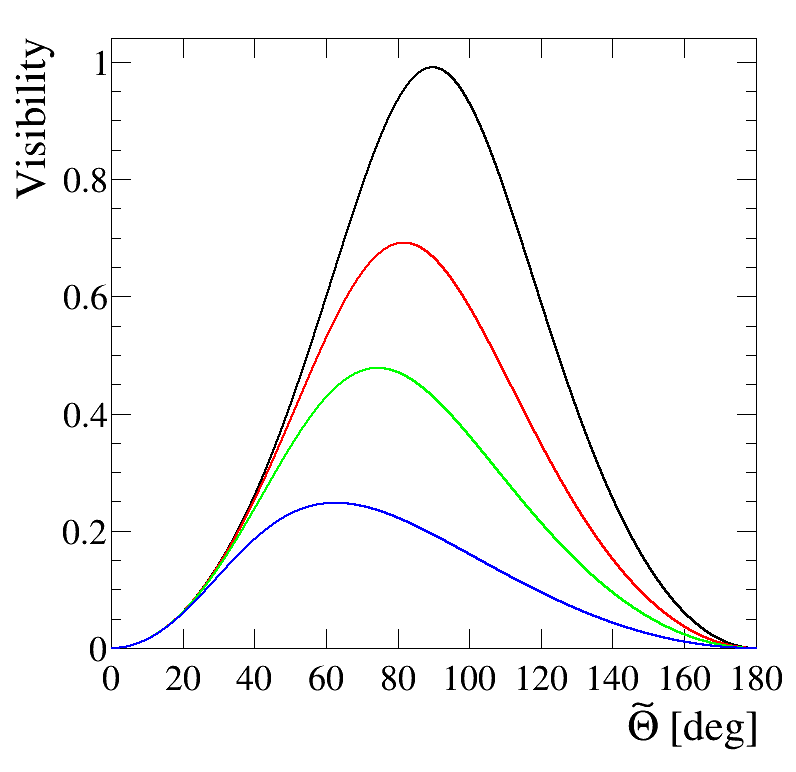}{}{fig:vis_env:vis}
{(a) The envelope function $\sF$ for the given incoming photon energy $k$ and the scattering angle $\tTheta$ defines the maximum amplitude of the visible oscillations for polarized photons.  The line colors represent different energies: black $k=0.1$, red $k=1 (511~\text{keV}) $, green $k=2$ and blue $k=5$. (b) The visibility or interference contrast is the component that reduces the oscillation patterns depending on the incoming photon energy $k$ and the scattering angle $\tTheta$. In particular, if the visibility is $0$, then no oscillation in $\Phi$ can be observed. This is the case for all energies if we consider small-angle scatterings or backwards scattering.}
{fig:vis_env}
{0.5}

To sum up, the scattering is ruled by an envelope function $\sF$, the probability for observing the polarisation depending on the Compton scattering angle and the initial energy of the incoming photon. The oscillation in the spatial scattering is ruled by $\Phi$ depending on the initial polarisation of the incoming photon. However, this information can only be extracted if and only if the visibility $\sV$ is non-zero (or high enough for experimental efficiency).

In particular, if we assume the source is producing an unpolarised state, i.e. $\rho_{HH}=\rho_{VV}=\frac{1}{2}$ and $\rho_{HV}=\rho_{VH}=0$, equation (\refEq{eq:cross-section-probability-single-photon}) predicts that no $\Phi$ dependence can be observed, whereas if the source produces e.g. a pure fully polarised state, e.g. $\kH$, it predicts an oscillatory behaviour with $\Phi$. Surprisingly at first glance, the formula also predicts a vanishing oscillation for a pure $|45\rangle$-polarized state. This is due to the summation over all final states, assuming the electrons in the plastic to be totally unpolarized.

In summary, the theory based on the Klein-Nishina formula predicts for $511$~keV photons a maximum visibility $\sV$ for a Compton scattering angle of $\tTheta_{\rm{max}}=81.66^\circ$, whereas no oscillation, i.e. no dependence on the polarisation, is observable if the $\sV$ is too small.

\subsection{The Klein-Nishina formula for multiphotonic Compton scattering events}
\label{subsec:klein-nishima-multi-photons}

The advantage of the pseudo-Kraus operator representation of the Klein-Nishina formula is its straightforward generalization to multi-photon states.
For $z$ photons we have\refPub{quantum1}
\CManyLinesEq{eq:cross-section-probability-many-photons}
{
 \DCS{\rho} = \left(\re\right)^{z} \sum_{\alpha=a}^{z} \left(\frac{k'_\alpha}{k_\alpha}\right)^{2} \sum_{l_a,l_b,\dots l_z=1}^{2}\\
 \TR{
  \left(\bigotimes_{\alpha=a}^{z}\Kr_{l_\alpha}^{(\alpha)}\right)\rho\left(\bigotimes_{\alpha=a}^{z}\Kr_{l_\alpha}^{(\alpha)\dagger}\right)
 }
}

Let us now focus on a source producing two photons in an arbitrary bipartite state $\rho$, which we can generally decompose into the Pauli matrices $\hat{\sigma}$ by
\CManyLinesEq{eq:rho-general}
{
\rho = \frac{1}{4}\biggl\{ \id_2\otimes\id_2+\vec{a}\cdot\hat{\vec{\sigma}}\otimes\id_2
+ \id_2\otimes\vec{b}\cdot\hat{\vec{\sigma}}\\+\sum_{i,j} c_{ij}\; \Pauli{i}\otimes\Pauli{j}\biggr\}
}
where $a_i,b_i,c_{ij}$ are real numbers. Considering only locally maximally mixed states, i.e. setting $a_i=b_i=0$, the remaining state space corresponds to a three-dimensional tetrahedron, a magic simplex~\refPub{MagicSimplex1,MagicSimplex2},  where the maximally entangled states are represented by vertices. All those states are equal in the property that there is only information encoded in the whole system, but not in any subsystems (individual photons). However, note that not all polarization states are possible if the bosonic nature of photons, i.e. that the total wave function has to be symmetric under particle exchange, is taken into account (for more details see Ref.~\refPub{quantum1}).

If we intend to describe pairs of photons distributed in different directions (but still opposite directions) our differential cross section is dependent on three pairs of coordinates $\theta, \phi$ (to describe the propagation direction of the two opposite moving particles), $\theta_a, \phi_a$ to describe the direction of propagation of the first scattered photon and $\theta_b, \phi_b$ of the second scattered photon. 

We should only choose one coordinate system to describe these two photons moving in opposite directions, also in the simulation. The Kraus operator $\mathcal{K}_2$ relates the propagation direction of the incoming photon (chosen by $\theta,\phi$ with respect to the laboratory frame) and the propagation direction of the outgoing photon $a$ (chosen by  $\theta_a,\phi_a$). Likewise, the second photon, moving in the opposite direction, is described by $\pi-\theta,\phi+\pi$  and the outgoing photon $b$ by  $\pi-\theta_b,\phi_b+\pi$. Only for one photon we can choose the propagation direction such that the Kraus operator gets independent of $\phi,\phi_{a}$ by choosing  $\phi=\phi_a$ or  $\phi=\phi_b$, i.e. the Kraus operator does then only depend on the Compton scattering angle $\tTheta_a$. In this case, the Kraus operator $\mathcal{K}_2$ of the second photon gets dependent on the scattered photon of the first photon, i.e. $\phi_a,\phi_b,\theta_b$, or differently stated also the Pauli matrices $\Pauli{1},\Pauli{2}$ give non-zero contributions. Note that this has nothing to do with the initial bipartite state being separable or entangled; it is solely due to the fact that we use one reference system to describe two particles. The description has also to be independent of the choice of which photon is first detected; however, in the simulation, we have to define which particle is $a$ or $b$ to derive the predictions of the theory defined by the formula, given in~\refEq{eq:cross-section-probability-many-photons}. Since the second Kraus operator expands whenever the two scattering planes are not the same to all Pauli matrices, one also has access to those components of the initial state~(\refEq{eq:rho-general})  via the scattering events.

\section{Interpretation of the Compton scattering as an estimator of polarisation}
\label{sec:experimental-determination-photon-polarization}

As aforementioned, for high-energetic photons such as e.g. $511$~keV, it is impossible to measure the polarisation using standard techniques elaborated in optical physics. But as outlined in the previous section, the polarisations degrees of freedom encoded in $\rho$ are observable via the oscillation if the visibility $\mathcal{V}$ is non-zero. 

If we assume that we have a source that produces the same linear polarized state, say $|H\rangle$, i.e. fixing a particular direction orthogonal to the propagation direction $\vec{k}$, then formula \refEq{eq:cross-section-probability-single-photon} tells us that the maximum value is exactly reached if $\Phi=\frac{\pi}{2}$ or $\Phi=\frac{3\pi}{2}$.  If we had a perfect polariser we would turn it with the angle $\Phi$ to observe the maxima and minima. In Compton scatterings, this may be identified with the observed scattering planes, i.e. the planes formed by the incoming $\vec{k}$ and outgoing $\vec{k}'$ trajectory of the photon, an experimentally accessible quantity in J-PET (see \refFig{fig:simple-compton-interaction}). Then we may interpret for a single photon event that Nature chooses for us $\tTheta,\Phi$, i.e. the quantization direction of the measurement device with respect to the incoming photon.  In other words, for fixed $\tTheta$ events our polariser changes randomly its direction; thus, if by chance the polariser was oriented at $\Phi=\frac{\pi}{2}$ we find the most events since it equals the polarisation of the source.

On the other hand, if we do not know the initial polarisation state, we can say that the measurement of a particular scattering plane, given by $\tTheta,\Phi$ does give us an estimate
of the polarisation of the incoming photon. Namely, we identify the direction perpendicular to the scattering plane as the polarisation direction of the incoming photon, i.e. $\hat{\epsilon} \equiv  \hat{k} \times \hat{k'}$, which information is given in the experiment by two hits in the scintillators, see \refFig{fig:simple-compton-interaction}. This estimator was put forward in Ref.~\refPub{quantum2} and is widely used by the J-PET collaboration.

However, our measurement device has an intrinsic uncertainty which can be estimated by the formula \refEq{eq:cross-section-probability-single-photon}. E.g. for incoming photon energies $k$ of $511$~keV, the fixed initial polarization state $\rho=|H\rangle\langle H|$ and the maximum $\tTheta_{\rm{max}}=81.66^\deg$,  we consider the normalized cross section, as a conditional probability density function:
\begin{eqnarray}
f(\Phi|\tTheta_{\rm{max}})&=&\frac{ \frac{d \sigma_\rho(\tTheta_{\rm{max}},\Phi)}{d\Omega}}{\int_0^{2\pi}\;\frac{d \sigma_\rho(\tTheta_{\rm{max}},\Phi)}{d\Omega}\; d\Phi}
\end{eqnarray}
Next, we compute the angle region around $\Phi_{max}=\frac{\pi}{2},\frac{3 \pi}{2}$ such that it covers $68\%$ of the events and we interpret the half of this region as one standard deviation.  This gives a value of about $42^\deg$. Therefore, if we considered a source of photons with the same $|H\rangle$ initial polarization state, then in $68\%$ of the cases the angle between the scattering planes and the initial polarization direction would lay in a range not wider than $[48^\deg, 132^\deg]$ or  $[228^\deg,312^\deg]$. 

\addOneFig
{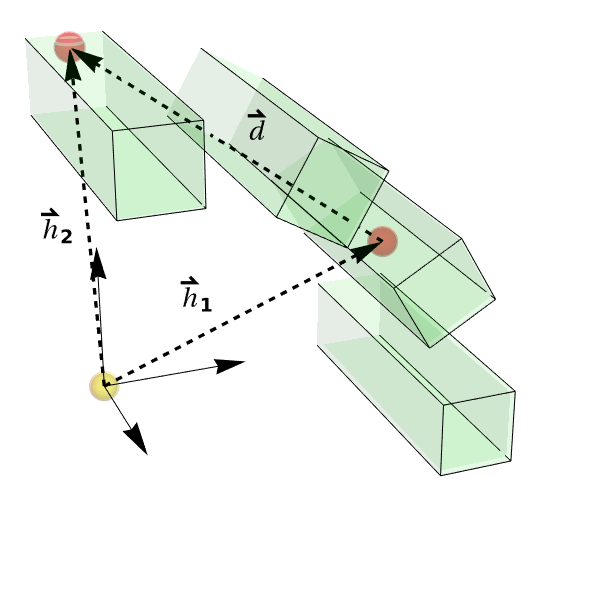}
{Photon momentum direction determination. The cuboids represent four detector modules, and the red spheres correspond to points of interaction via Compton scattering. The momentum directions of the photon before first Compton scattering is calculated as $\hat{k}=\frac{\vec{h}_1}{|\vec{h}_1|}$ and before the second scattering as $\hat{k'}=\frac{\vec{d}}{|\vec{d}|}$ where $\vec{d}=\vec{h}_2-\vec{h}_1$. The calculations assume that the photon source (yellow sphere) is at the center of the coordinate system.}
{fig:simple-compton-interaction}{0.5}

In the case of two photons, typically one has experimentally access to the difference between the maximum and the minimum of the recorded events in dependence of the difference of the azimuthal angle $\DeltaPhi$ as defined in \refFig{fig:scattering-planes}, which must be related to the polarisation of the initial two-particle state.  Following up the considerations from the single photon case where we have interpreted the direction normal to the scattering plane, $\vec{n}_i = \hat{k}_i \times \hat{k'}_i$, as the polarisation of the incoming photon $i$, consequently the angle between the two scattering planes can be related to the difference in their polarization, i.e.
\OneLineEq{eq:scattering-plane-vector}
{\DeltaPhi \equiv \phi_1 - \phi_2 \equiv \angle (\vec{n}_1,\vec{n}_2)\equiv \angle (\hat{\varepsilon}_1,\hat{\varepsilon}_2) \;.}
If we assume that $\theta_a,\phi_a$ and $\theta_b,\phi_b$ are chosen both independently, then the uncertainties in the polarization determination of the single photon case multiply accordingly.

\addOneFig
{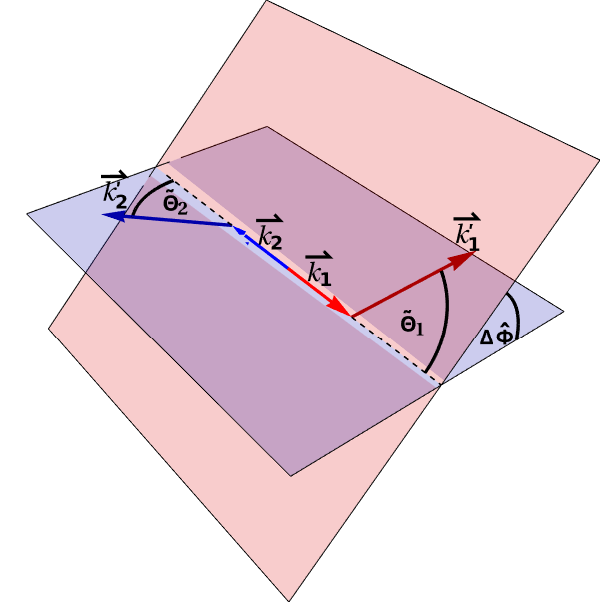}
{Estimation of the relative angle between the scattering planes of the photon pair. The vectors $\vec{k}_{1/2}$ show the directions of the annihilation gamma's initial momentum. The vectors $\vec{k'}_{1/2}$ are the momentum directions after the Compton interaction with the detector medium (scintillators). The yellow and blue planes represent the scattering plane of the first and second gamma, respectively. The relative angle $\DeltaPhi$ is observable that relates to the initial joint polarisation state of the photon pair (see ~\refEq{eq:cross-section-probability-many-photons}).}
{fig:scattering-planes}
{0.5}

\section{`\textit{Vienna-Warsaw Simulation Model}' capable of simulating any initial bipartite polarisation state}
\label{sec:simulation-model}

There exist sources that produce a two-particle initial state, which may be separable or entangled (and pure or mixed). Indeed, it is an open question, what initial states have to be assumed for e.g. of photons produced by direct annihilation processes or via the decay of para-positronium or via the decay of ortho-positronium via a pick-off or spin exchange processes~\cite{bass_colloquium_nodate}. In practice one often has to deal with a mixture of all those processes. Currently, researchers are also working on generating positronium beams with defined quantum properties.

The Geant4 Monte Carlo simulation toolkit is an open-source software package dedicated to numerical simulations used extensively in nuclear physics, particle physics, astrophysics and medical physics studies~\refPub{allison2016}. 
It covers the modelling of decay kinematics as well as photon interactions with the detector material.  
Several different implementations of the underlying electromagnetic physics processes can be chosen in the Geant4 simulation. The Livermore low energy model\refPub{depaola_new_2003,agostinelli_geant4simulation_2003} includes the Compton scattering processes description,  which takes into account the polarisation degree of freedom of an individual photon, more precisely, its linear polarisation degree can be chosen to be orthogonal. 
However, as discussed later in detail, the Livermore model describes correctly only the individual linear polarisation states of each of the two photons~\footnote{Starting from the GEANT4v11.0, the G4LivermorePolarizedComptonModel and G4eplusAnnihilation models were extended to {\it entanglement mode}. This model would correspond to the simulations of the $\Psi^{+}$ function, as described in~\refPub{watts_photon_2021}.}. In particular, it does not allow the simulation of arbitrary joint states of photons.

Therefore, we developed a dedicated  VW simulation model, based on the formalism presented in the previous section, and incorporated it into the Geant4 framework (version 10.05.p01).
VW model provides the possibility for simulating Compton scattering kinematics for pairs of photons with an arbitrary initial joint polarization state defined by simulation configuration options. Moreover, the program provides options for averaging over initial and/or final polarization states. 
Functionally, the implementation consists of two components (see \refFig{fig:simulator_scheme}). The Model Factory is responsible for the generation of the differential cross-section describing the double Compton scattering process based on the 
initial two-photon polarization state. A general polarization state can be defined by a set of parameters provided by the user. Alternatively, several predefined polarization states are available.  Next, the Model Factory generates the corresponding differential-cross sections or uses one of the cached functions for predefined polarization states. 
The cross-sections are subsequently used in the Geant4 simulation workflow for the photon pairs marked as correlated, as shown in the diagram \refFig{fig:simulator_scheme3}. The idea of the correlation is implemented in the following way: The angular and polarization states are calculated for both photons when the first interaction occurs. It is applied to the first photon and saved for future usage. When the second photon interacts, the precomputed values are used. In case of consecutive interactions, the workflow switches back to the standard Geant4 Livermore library.   

\addTwoFigs
{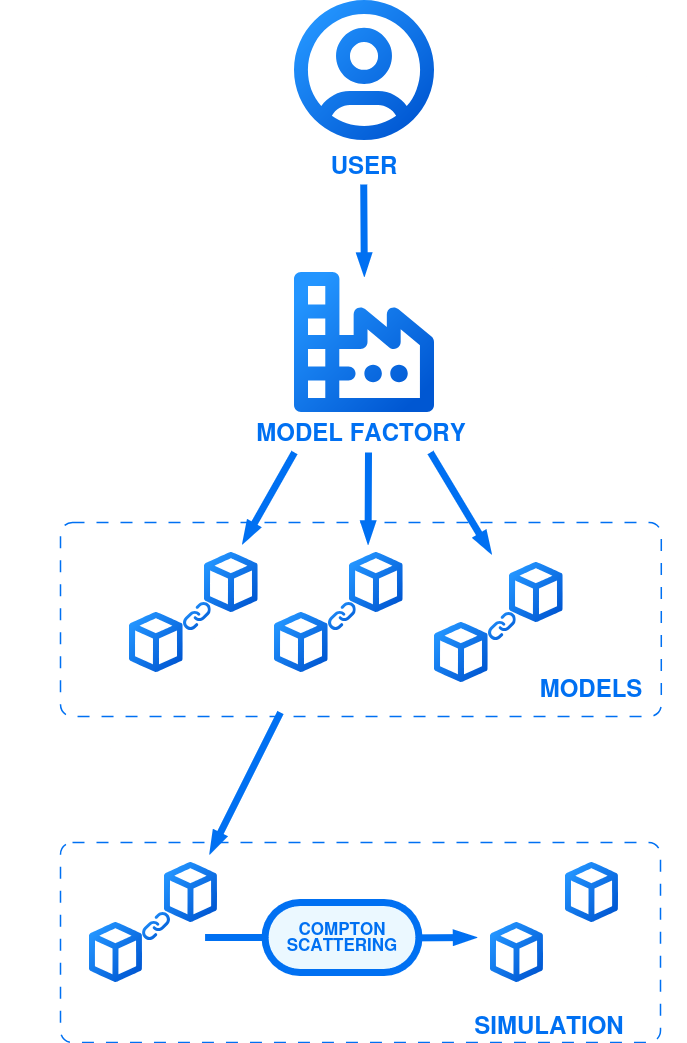}{The initial two-photon polarization state is set based on the parameters provided by the user. Next, the Model Factory calculates the corresponding double differential cross-section or uses one of the cached functions for chosen polarization states. The differential cross-section is used in the Geant4 simulation workflow.}{fig:simulator_scheme1}
{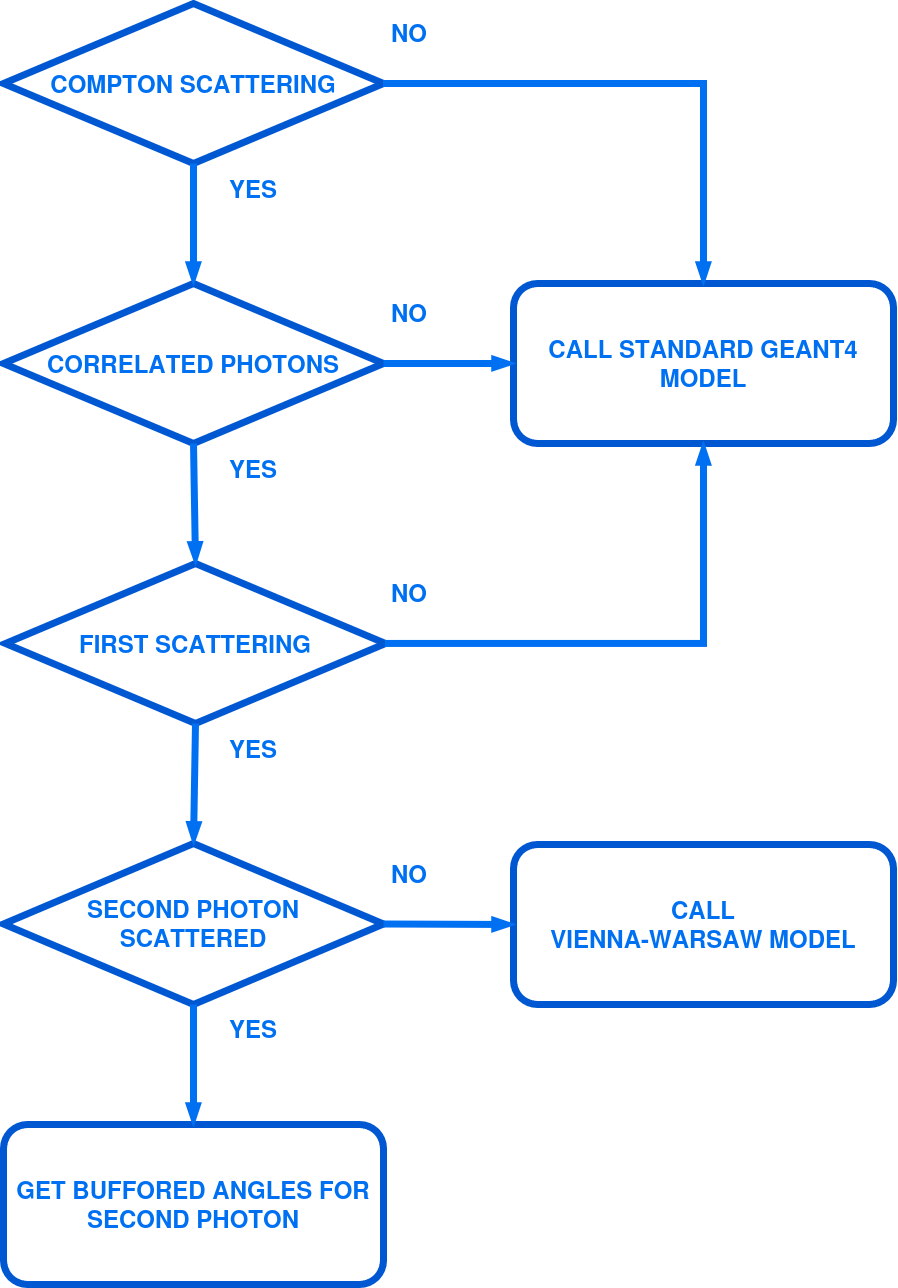}{Geant4 simulations workflow. For the correlated photons, the respective cross-section generated by the VW model is applied.}{fig:simulator_scheme3}
{Simulator scheme}
{fig:simulator_scheme}
{0.4}

To ensure the self-consistency of the VW model implementation, we developed several automated tests. We reproduced the shapes of the visibility and envelope functions introduced in \refSec{sec:quantum-model}. For various two-photon polarization states e.g. all Bell states, we numerically cross-checked the generated cross-section values, for a large range of $\theta$ and $\phi$, with the theoretical predictions. Also, we tested that our VM simulator reproduces numerically the theoretical relation between the mutually unbiased bases (MUB)  and symmetric informationally complete positive operator valued measures (SIC-POVMs) (for more details see Ref.~\refPub{quantum1}).

\subsection{VW Model applied to different initial bipartite states for ideal detectors}

\addThreeFigsH
{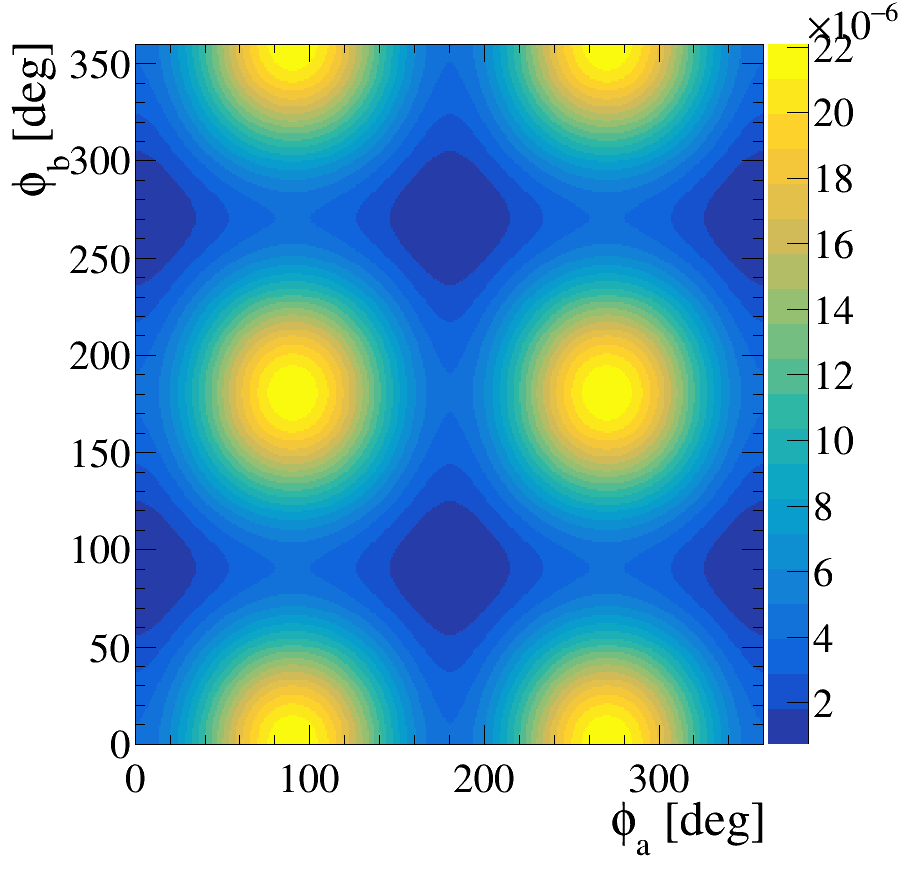}{}{fig:phi_a_vs_phi_b_hv_and_vh:hv}
{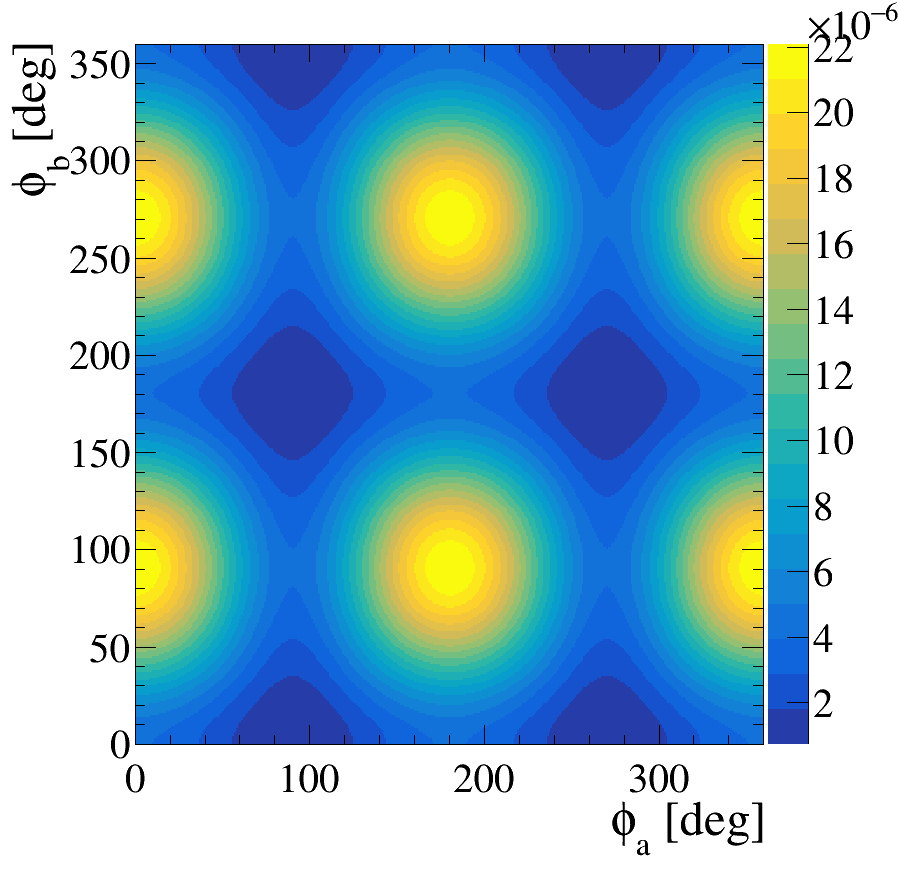}{}{fig:phi_a_vs_phi_b_hv_and_vh:vh}
{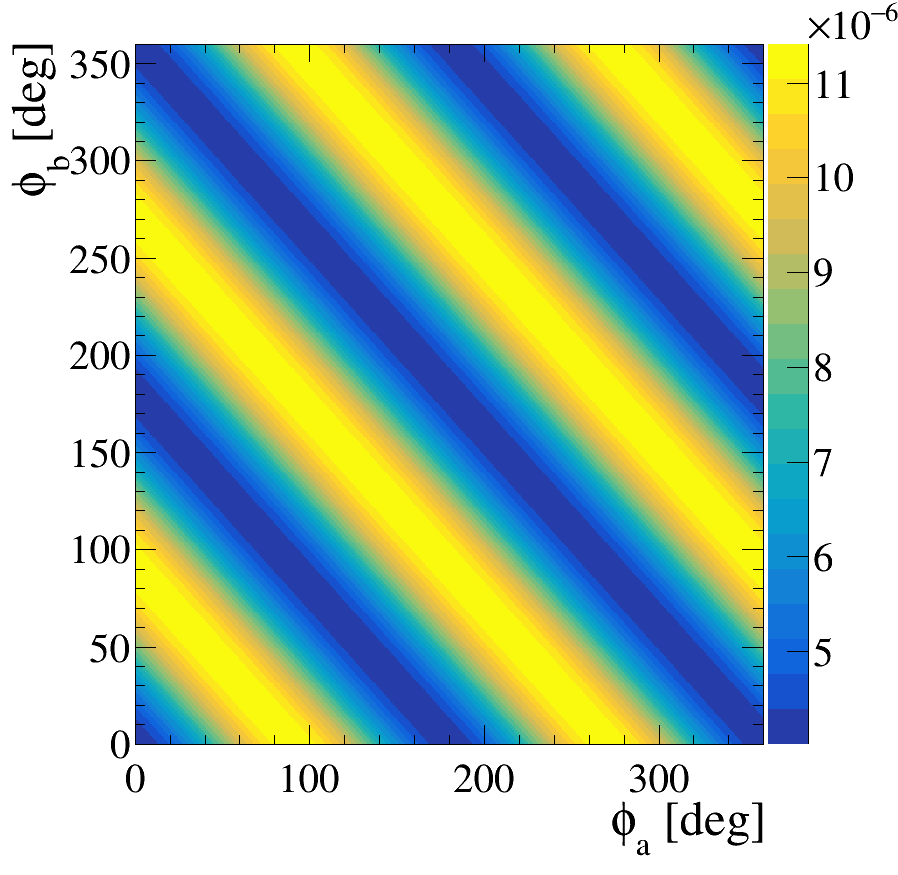}{}{fig:phi_a_vs_phi_b_hvvh:psi-plus:rho}
{The distributions of azimuthal angles $\phi_a$ versus $\phi_b$ of primary photon pairs under the assumption of perfect detection efficiency for four different initial quantum states.  The difference between the histograms in (a,\refEq{eq:HVstate}) and (b,\refEq{eq:VHstate}) is a shift of $90^\deg$ corresponding to the definition of $\ket{H}$ and $\ket{V}$ assuming the classic mechanics model. Histogram (c,\refEq{eq:rhomixed} and \refEq{eq:state-psi0pm}) can be obtained in two ways: as a statistical mixture of state $\ket{HV}$ and $\ket{VH}$ or for the entangled state $\ket{\psi^{+}}$ - in both cases, we assumed the quantum mechanics model. We assumed here that both annihilation photons propagate along the $z$-axis.}{fig:phi_a_vs_phi_b_hv_and_vh}{0.21}

To show the power of VW model and its agreement with the predictions of the theory, let us start with a simplified, idealized experiment. We assume that a point source is placed in the centre between two detectors and emits a pair of photons, which, due to momentum and energy conservation, propagate in opposite directions in e.g. $z$-direction, where we locate our two detectors. Furthermore, we place another detector at some Compton scattering angle for each photon to register the scattered photons. Thus, we have a four-hit event, which mimics typical setups for optical photons. The crucial difference is that the angle $\DeltaPhi$ is the observable that relates to the initial polarisation state of the two photons, however, due to the averaging over the final polarisation degrees of freedom due to the detector material and the dependence on the energy and scattered angle it does only give partial information about the initial polarization.

For our idealised setup, let us assume $100\%$ efficiency in detecting the photons, so the intensity should vary with $\DeltaPhi=\phi_a-\phi_b$.

In \refFig{fig:phi_a_vs_phi_b_hv_and_vh} we plot the results of the VW model assuming as an initial state that would correspond in quantum theory to  (a) \begin{equation}\label{eq:HVstate}
  \ket{\Phi_1} = \kH\otimes\kV=\kHV  
\end{equation} and (b) 
\begin{equation}\label{eq:VHstate}
  \ket{\Phi_2} = \kV\otimes\kH=\kVH  
\end{equation}
and (c) 
\begin{eqnarray}\label{eq:rhomixed}
\rho_{mixed} &=& \frac{1}{2} \ket{\Phi_1}\bra{\Phi_1}+\frac{1}{2} \ket{\Phi_2}\bra{\Phi_2}\nonumber\\&=&\frac{1}{2}\kHV\bHV+\frac{1}{2}\kVH\bHV
\end{eqnarray}
and  (c')
\OneLineEq{eq:state-psi0pm}{\ket{\psi^+} = \frac{1}{\sqrt{2}}\big\lbrace\ket{\Phi_1}+\ket{\Phi_2}\big\rbrace=\frac{1}{\sqrt{2} }\big\lbrace\kHV+\kVH\big\rbrace\;.}
Obviously, the first three states are separable with respect to the Hilbert space of the two photons $\mathcal{C}_2\otimes\mathcal{C}_2$, whereas the last one is a maximally entangled state, a so-called Bell state. Furthermore, note that the last two states have in common that the subsystem states, i.e. the individual photons, are totally maximally mixed, a particular property exhibited by all convex combinations of maximally entangled Bell states which form a so-called magic simplex in the Hilbert-Schmidt space~\refPub{MagicSimplex1,MagicSimplex2}. Important for our studies here that no correlations are considered for the single photons, i.e. the states correspond to the set of locally maximally mixed states.

Furthermore, except for the third state $\rho_{mixed}$ complete information of the initial state is available since the states are pure, whereas $\rho_{mixed}$ includes the classical uncertainty of not knowing which of the two states the source has been sent. 

 Consequently, adding the weighted sum of the two individual simulations (\refFig{fig:phi_a_vs_phi_b_hv_and_vh:hv}/ \refFig{fig:phi_a_vs_phi_b_hv_and_vh:vh}) leads to \refFig{fig:phi_a_vs_phi_b_hvvh:psi-plus:rho}. However, assuming $|\psi^+\rangle$ leads also to \refFig{fig:phi_a_vs_phi_b_hvvh:psi-plus:rho}, but here each event has the same individual statistics behind each of both cases $\ket{HV}$ and $\ket{VH}$ in the simulation, in strong contrast to the previous case. One observes the strips of destructive and constructive interference.

\subsection{Comparison of `\textit{VW Simulation Model}' with current simulators}

Let us compare the VW simulation model to other available models. The Geant4 physics list allows choosing the \textit{Standard Livermore Model}, which implements the Compton scattering events without taking the polarization degrees of freedom into account. Obviously, in this case, no oscillatory shapes in the $\DeltaPhi$ distribution are expected.  A similarly flat distribution is also predicted for the maximally mixed state $\frac{1}{4}\mathbbm{1}_2\otimes\mathbbm{1}_2$. Indeed, as shown in ~\refFig{fig:comp_QE_lpm_lm_2hits}. the VW model correctly generates the flat line for this case.  
\addTwoFigsH
{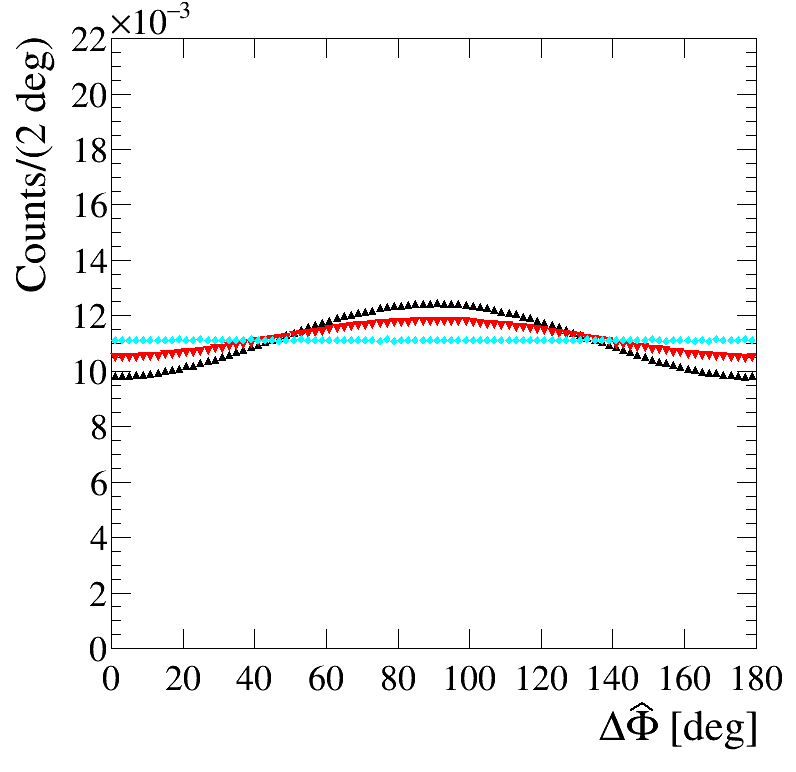}{}{fig:comp_QE_lpm_lm_2hits:no-cut}
{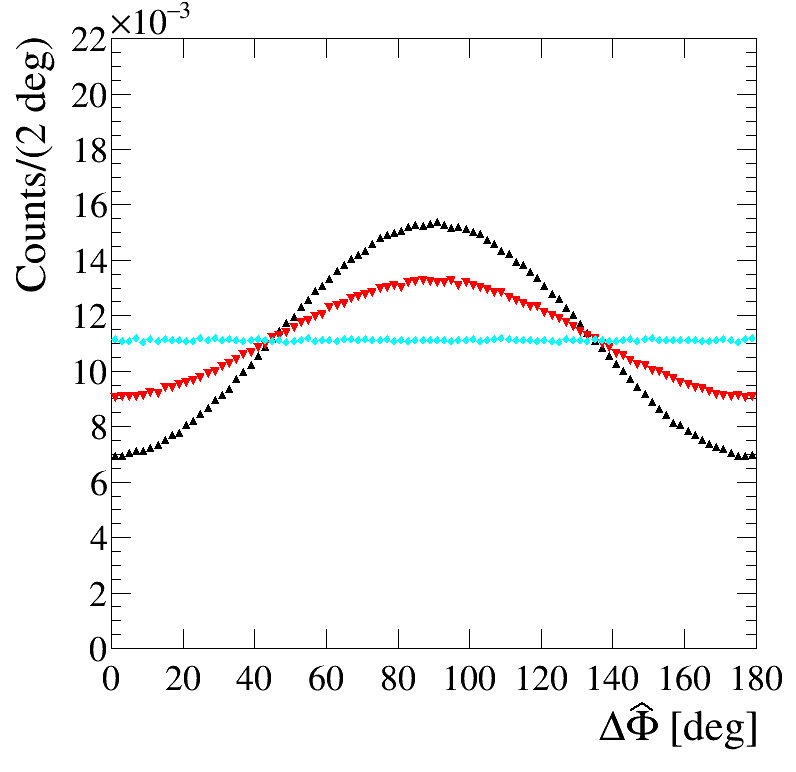}{}{fig:comp_QE_lpm_lm_2hits:circle-cut}
{ Distribution of the $\DeltaPhi$ observable generated by the VW model for (a) all Compton scattering angles and (b) restricted scattering angles to the circle area with the radius of $30^\deg$ and the centre placed at the maximal visibility angles for 511 keV photons $\tTheta_{\rm{max}}=\tTheta_a=\tTheta_b=81.66^\deg $. The (black triangles) graph corresponds to the result assuming an initial maximally entangled state  $\ket{\psi^+}$, the (red down triangles) graph assumes the classical mechanics model of polarised photons, and the  (blue diamonds) graph corresponds to the initial unpolarized state. The last two scenarios agree with the results of the  Polarized Livermore  Model and Standard Livermore Model, respectively. Restricting the Compton angle region leads to higher visibility but lowers the overall statistics.}
{fig:comp_QE_lpm_lm_2hits}
{0.5}
In the \textit{Livermore Polarized Model} the linear polarization of the individual photons can be chosen to be orthogonal to each other. This effectively leads to the $\DeltaPhi$ distribution which would correspond to the simulation of a separable quantum state of type $\kHV$.
Indeed, the same shape is reproduced by the VW model with the polarization state set to $\kHV$ (see \refFig{fig:comp_QE_lpm_lm_2hits} red down triangles.).

\section{Experimental data analysis}
\label{section:experimental-data}

We first give a very short introduction to the detection system and the source, followed by showing the experimental data set.
 
\subsection{Detection system and positronium source }

The study is performed based on a subsample of the data collected in 2020 by the J-PET Big Barrel detector\refPub{niedzwiecki_j-pet_2017,dulskiJPETDetectorTool2021a}. 
It consists of three scintillator layers oriented cylindrically. Each scintillator has $19$~mm $x$-length, $7$~mm $y$-length and $500$~mm $z$-length. The first layer has $48$ scintillators and a radius of $425$~mm, the second layer has $48$ scintillators shifted relative to the first layer with $3.75^\deg$ and a radius of $467.5$~mm, the third layer has $96$ scintillators shifted relative to the first layer with $1.875^\deg$ and a radius of $575$~mm.
The detector readout system consists of the Hamamatsu R9800 vacuum photomultipliers connected from both ends to each EJ-230 scintillator strip, which transforms the generated light into electric signals. The signals are probed at four thresholds and stored by the dedicated triggerless data acquisition system (DAQ)\refPub{korcyl}.      
The spatial resolution along the scintillator strip is equal to $\sigma=2.5~\text{cm}$, while the Time-of-Flight resolution corresponds to about 220 ps~\refPub{niedzwiecki_j-pet_2017}.

A ${^{22}\mbox{Na}}$ radioisotope is used as a source of positrons. The atom of ${^{22}\mbox{Na}}$ decays via $\beta^{+}$process: 
${^{22}\mbox{Na}} \rightarrow {^{22}\mbox{Ne}}^{\star} + e^{+} + \nu_{e} $   emitting a positron which after thermalization is likely to interact with an electron and form a positronium atom, which subsequently decays into photons. The ${^{22}\mbox{Ne}}^{\star}$ isotope deexcites by emitting the prompt gamma with an energy of 1275 keV. All photons can be detected by the J-PET detection system via the Compton scattering process.
A 0.7 MBq activity source in the form of a micro-droplet of liquid ${^{22}\mbox{NaCl}}$ is packed in the thin Kapton foil and sandwiched in the layers of 3 mm thickness XAD-4 porous polymer, which enhances the production of the ortho-positronium atoms \refPub{gorgol_construction_2020}. The source is placed inside the vacuum chamber and put in the geometrical centre of the scanner.

\subsection{Data selection}
The collected sample was preprocessed using the J-PET Framework software\refPub{jpetFW}. 
In the preprocessing phase, several reconstruction steps are performed, starting from the reconstruction of the electric signal from the photomultiplier, through the hit position corresponding to the photon Compton interaction position in the scintillator module, till the events are formed as sets of hits connected as candidates coming from the same radioisotope decay.       
The photon hit is defined based on photon energy deposition measured through time-over-thresholds (TOTs) of pairs of photomultiplier signals recorded at two ends of a scintillator strip for every registered photon interaction~\refPub{sharma_estimating_2020}. 
The formation of the event candidates is based on the selection of the four hits registered in separate scintillator strips within the coincidence time window of 20 ns.    

In addition, a pair of hits must be identified as {\it annihilation hits}, based on the set of conditions. First, the corresponding TOT values must lay within the range, denoted as red area in~\refFig{fig:experiment-cuts:tot}, panel right, close to the $511$-keV photon Compton edge. Next, the pair must fulfil a set of geometrical requirements based on the registered time difference and hit positions that limit the reconstructed annihilation point to the vicinity of the positronium source position. For more details see~\ref{appendix:annihilation_selection}. 
\addTwoFigsH
{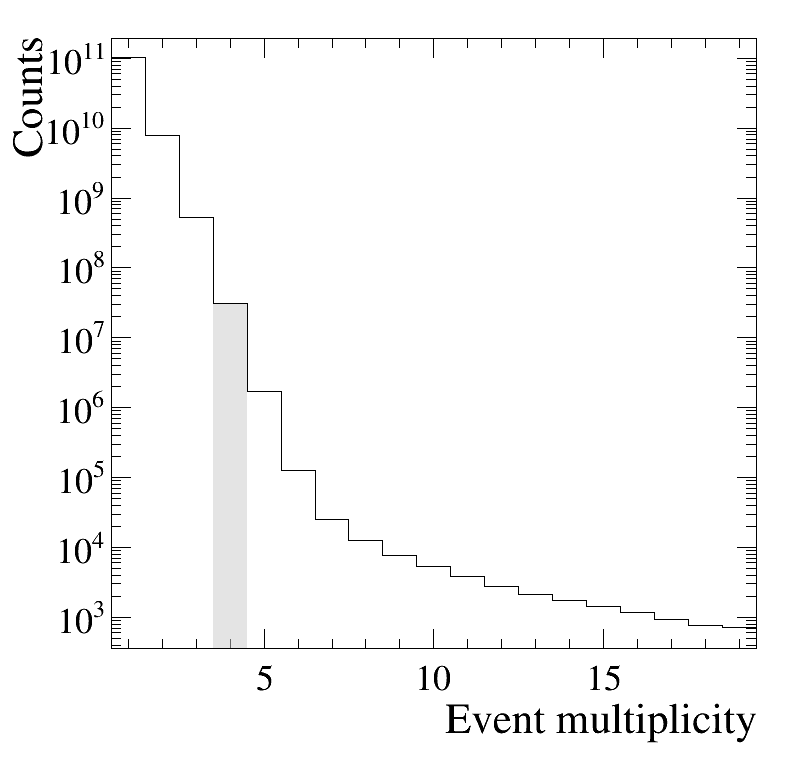}{}{fig:experiment-cuts:multiplicity}
{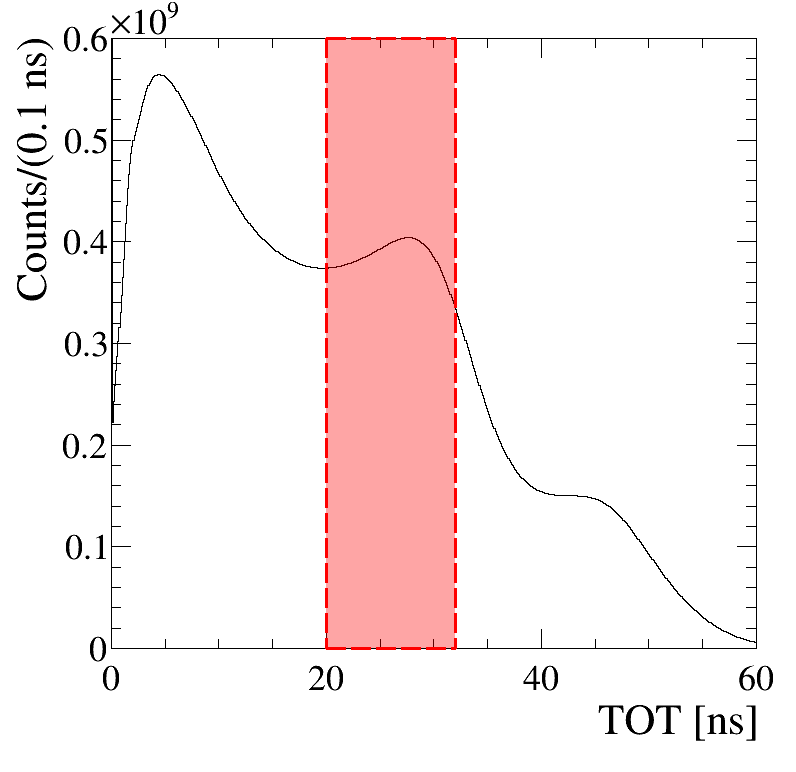}{}{fig:experiment-cuts:tot}
{(a) Distribution of the hits multiplicity per event candidate. The total number of events equals $1.5\cdot 10^{11}$, out of which about $3.2\cdot 10^{7}$ events have 4 or more hits. The grey rectangle represents the condition applied in the analysis.
(b) Experimental TOT distribution. The TOT values of 33 ns and 48 ns correspond to the Compton edges for the primary 511 keV annihilation photons and 1275 deexcitation photons, respectively. The red area between 20 ns and 32 ns corresponds to the annihilation hits assigned to the primary interaction in the detector.}
{fig:experiment-cuts}{0.45}

Out of the analysed events, $5.37\cdot 10^{5} $  events fulfilled the four-hit conditions. 
Next, pairs of hits are classified as corresponding to primary or secondary scattering interactions by comparing the registered time difference $t_{i} - t_{j}$ against the distance between hits' positions $d_{ij}$. For the two consecutive interactions of the same photon, the registration time difference corresponds to the travel time between the first and the second scattering, and it must be equal, within the experimental uncertainties,  to the distance between hits:  $ |t_{i} - t_{j}| \times c \approx  d_{ij}$. Therefore, the metric value  $\Delta_{ij} = |t_{i} - t_{j}| -   \frac{d_{ij}}{c}$ is used to find associated pairs of primary and secondary hits among the selected four hit events.   
The distributions of $\Delta_{1i}~\text{versus}~\Delta_{2i}$ for experimental data and for MC sample  are presented in~\refFig{fig:dvd-histograms:experiment} and ~\refFig{fig:dvd-histograms:simulation}, respectively. After the application of all selection criteria, the directions of the photon momenta are determined based on the primary and secondary hit positions as described in \refFig{fig:simple-compton-interaction}. 
Finally, for each event, the $\DeltaPhi$ value is calculated.

\addTwoFigsH
{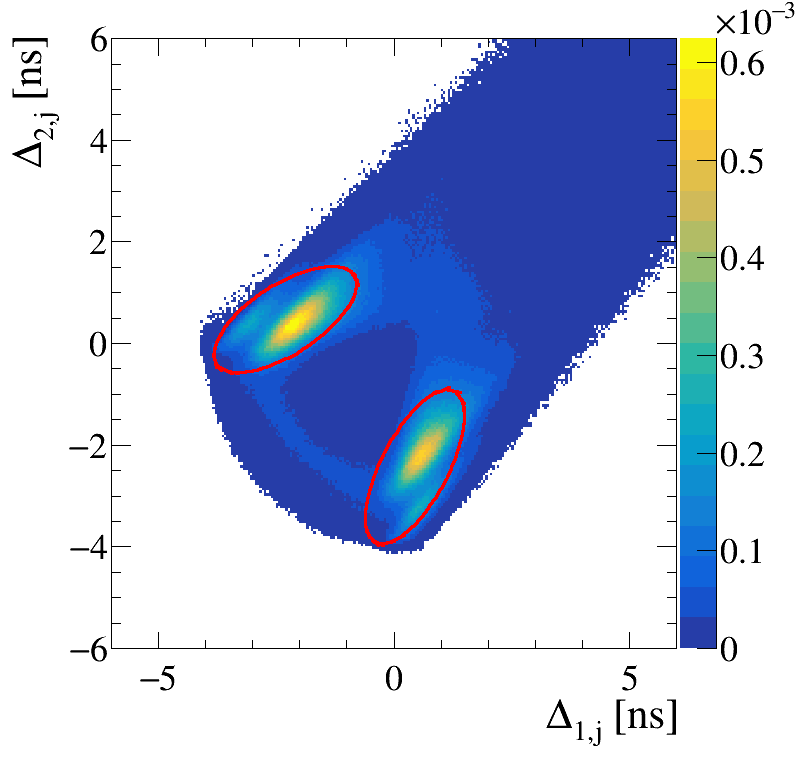}{Experimental data}{fig:dvd-histograms:experiment}
{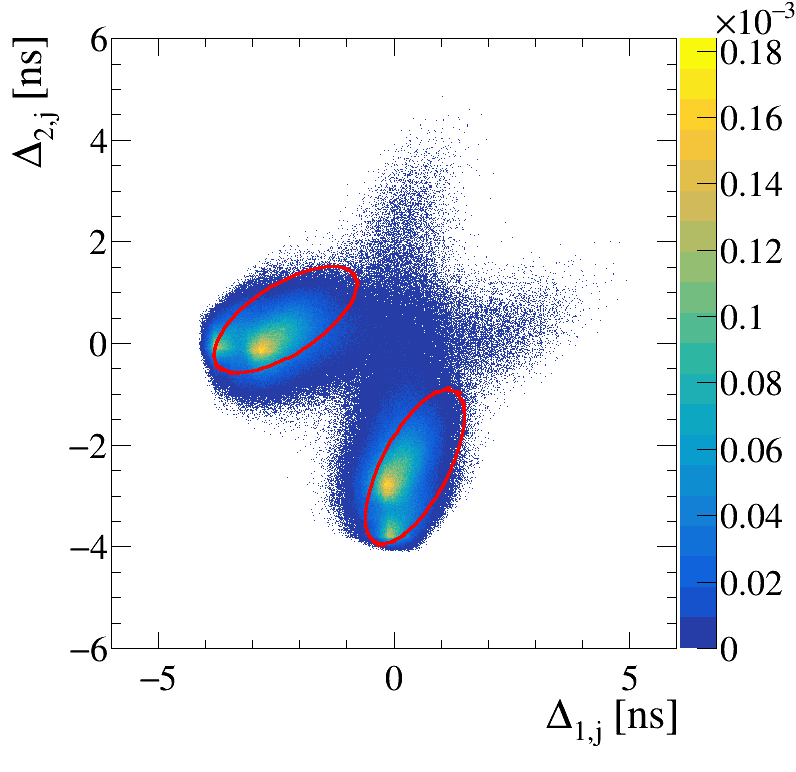}{MC simulations}{fig:dvd-histograms:simulation}
{Distribution of $\Delta_{1i}~\text{versus}~\Delta_{2i}$ parameters,   expressing the difference of the spatial distance and the travel time difference $\Delta_{ij} = |t_{i} - t_{j}| -   \frac{d_{ij}}{c}$ between two registered interactions determined for the primary hits. The red ellipses correspond to the selection areas for pairs of interactions assumed as consecutive primary and secondary Compton scatterings originating from the same photon.  }
{fig:dvd-histograms}{0.5}

\subsection{MC samples}

The simulation models introduced in section~\refSec{sec:simulation-model} were used to produce the MC samples for a simulation of an initial quantum polarization state in two different basis choices: $$\ket{\psi^{+}(H/V)} = \frac{1}{\sqrt{2}} (\kHV +\kVH)$$
linear polarization basis $H/V$ basis and 
$$\ket{\psi^+ (\pm 45^\circ)} = \frac{1}{\sqrt{2}}\left(\ket{+45^\circ-45^\circ}+\ket{-45^\circ+45^\circ}\right)$$ $45^\circ$ rotated to the linear $H/V$ basis. 
Both states are equivalent in the quantum theory.

The simulated MC samples consisted of $2\cdot 10^{12}$  photon pairs, generated using the experimental setup corresponding to the J-PET Big Barrel detector. 
During the post-processing, the registered positions, times and energies were smeared using the phenomenological parameterizations modelling the experimental resolutions of the front-end electronics and the DAQ system. The energy resolution dependence is parameterised as a 
$\frac{\sigma(E)}{E} =  \frac{0.06}{\sqrt{\Delta E[MeV]}}$ 
fraction.
The simulated photon registration time is smeared, event by event, by replacing the event registration time $t_{r}$ by the value obtained from the normal distribution $N(t_{r},\sigma_{t}$), where 
$\sigma_{t} = 0.22\sqrt{\frac{340}{\Delta E[keV]}}~\text{ns}$ corresponds to the temporal resolution. 
Analogically, the registration position along the scintillator strips ($z$ position) is smeared, event by event, by replacing the registered photon position $z$  by the value obtained from the normal distribution $N(z,\sigma_{z})$, where $\sigma_{z} =2.5~\text{cm}$ corresponds to the positional uncertainty along the scintillator strip of the scanner.
Due to limitations of the front-end electronics, some events would involve 511 keV photons undergoing multiple interactions within the same scintillator strip. In such cases, all interactions can be recorded as a single signal. This effect is partially accounted for in our simulations by merging multiple scattering within the experimental resolution. However, it should be emphasised that it remains unclear whether the Geant4 LivermorePolarized model accurately describes the final polarisation state of photons after multiple scatterings in this context.  A more precise treatment would require the implementation of a dedicated multiple scattering model as proposed, for example, in~\cite{hiesmayrQuantumErrorChannels2024,caradonnaKinematicAnalysisMultiple2024}.
To mimic the measurement conditions and the TOT selection, we accepted only the events which consisted of four hits with the deposited energy lying in the range from 70 keV to 340.6 keV.  
The sample was subsequently filtered by applying the same selection procedure as in the case of the experimental data.
Out of all the generated events, the $4.18 \times 10^7 $  candidates fulfilled 4-hit conditions, and about $2.24 \times 10^5$ remained after applying the final selection procedure.

\subsection{Selection of high visibility region and detection efficiency corrections}
To enhance the correlation effects,  
we restrict the data to the area of the highest visibility by requiring the condition of the combination of the primary scattering angles. 
We limit the Compton scattering angles for first and second annihilation photons to lay within a circle area, defined by the centre corresponding  to the maximum visibility ($\tTheta_{\rm{max}}$) and the radius of  $R_{\tTheta_{\rm{max}}} = 30^\deg$, i.e.
\OneLineEq{eq:circle-theta-cut}{(\tTheta_1-\tTheta_{\rm{max}})^2 + (\tTheta_2-\tTheta_{\rm{max}})^2 \leq {R^2_{\tTheta_{\rm{max}}}}}
The selected area on the primary scattering angle distributions for the theory, the experimental sample and the MC simulation are shown in~\refFig{fig:experimetaldata:tvt}. By comparing the theoretical and experimental distributions, one can see that the J-PET detector has the highest acceptance concentrated in the region close to the maximum visibility point, e.g. the maximum event density can be found for  $(\tTheta_1,\tTheta_2) \approx (94^\deg, 94^\deg)$.

The selection of the high visibility region reduces the experimental sample to $2.5 \cdot 10^5$ event candidates.
\addThreeFigsH
{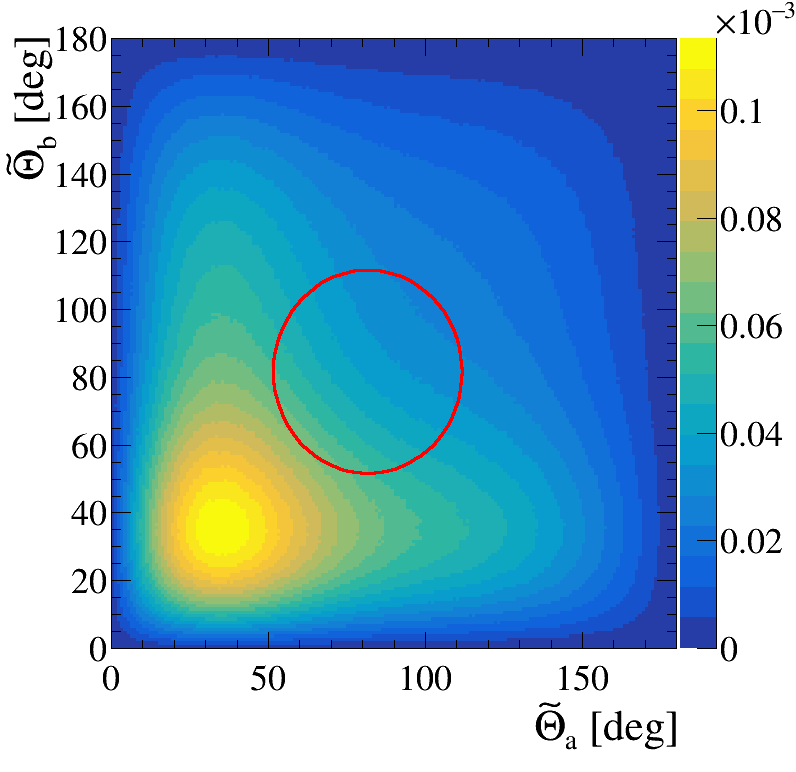}{Perfect detector}{fig:theta-vs-theta:ideal}
{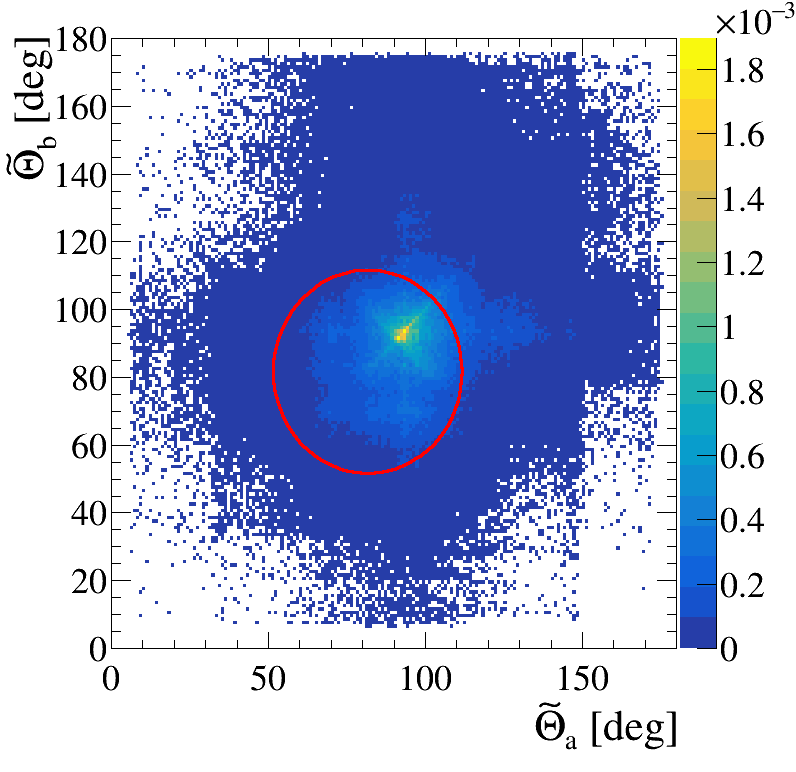}{Experimental data}{fig:experimetaldata:tvt:measurements}
{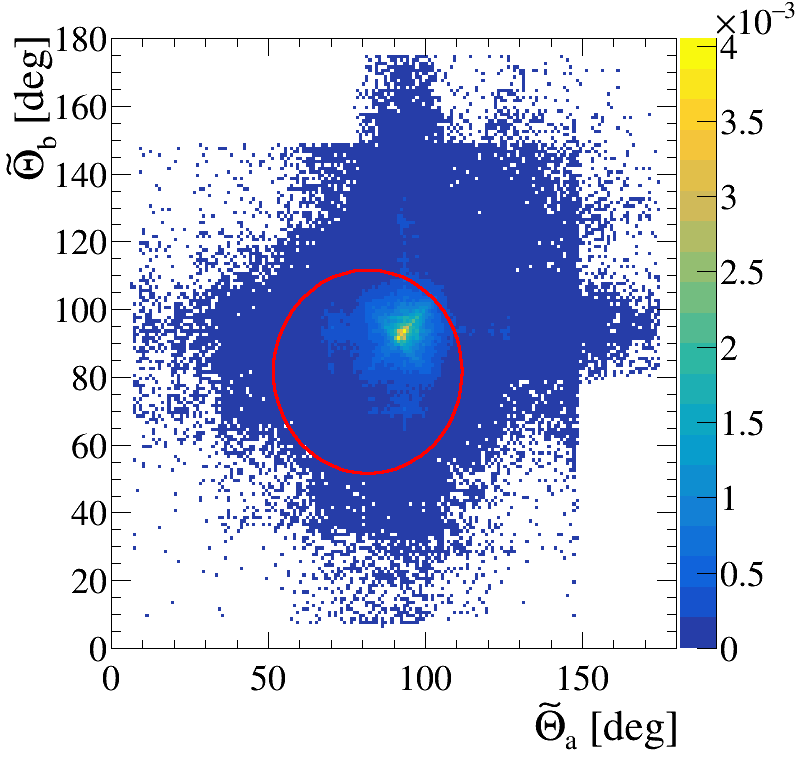}{MC simulations }{fig:experimetaldata:tvt:simulations}
{The scattering angle distributions  $\tilde{\Theta}_a~\text{vs}~\tilde{\Theta}_b$ of primary photons for (a) a perfect detector with 100\% geometrical and detection efficiency, (b) the experimental data and (c) MC simulations of   $\ket{\psi^{+}}  = \frac{1}{\sqrt{2}} (\kHV +\kVH)$ state.    The red circles represent the high-visibility area.}
{fig:experimetaldata:tvt}{0.21}

As aforementioned, the shape of the $\DeltaPhi$ distribution for the  $\ket{\psi^+(\pm 45^\deg)}$ initial state should be flat, and any systematic deviation from the flat line must be the manifestation of the detector-induced effects. Therefore, we use this MC sample to calculate the bin-per-bin  correction coefficients $\alpha_i$ that would take into account the detector-induced deformation in the $\DeltaPhi$ degree of freedom:
\OneLineEq{eq:correction-coefficient}{\alpha_i = \frac{\mu}{H_i}\qquad\textrm{with}\qquad\mu = \frac{1}{N}\sum_{i=1}^{N} H_i}
where $H_i$ is the number of registered events in the $i$-th bin of the $\DeltaPhi$ histogram.

\subsection{Correlation plots}

The shape of the $\DeltaPhi$ spectra can be described by $f(\DeltaPhi) = a(1 + b\,\cos(2\DeltaPhi))$ function, where $a$ represents the size of the quantum--uncorrelated part, while the  $\Vis^2=|b|$ can be interpreted as a squared visibility factor introduced in ~\refEq{eq:visibility}. The experimental data and MC simulations distributions are presented in \refFig{fig:correction-r1-fits}. Both distributions are corrected for the detector-induced effects. 
The shapes of both data and MC distributions are in very good agreement.
The weighted maximum likelihood technique was used to fit the $f(\DeltaPhi)$ distributions.
The extracted fit parameters for MC and experimental data are equal within the obtained uncertainties. 
The determined value of squared visibility is equal to  $\Vis^2_{exp}=0.27 \pm 0.02$ and $\Vis^2_{MC}=0.28 \pm 0.03$, for the experimental data and MC simulations, respectively. 
The quoted uncertainties are calculated based on the fitted parameter uncertainties and include only the statistical effects.

\addTwoFigsH
{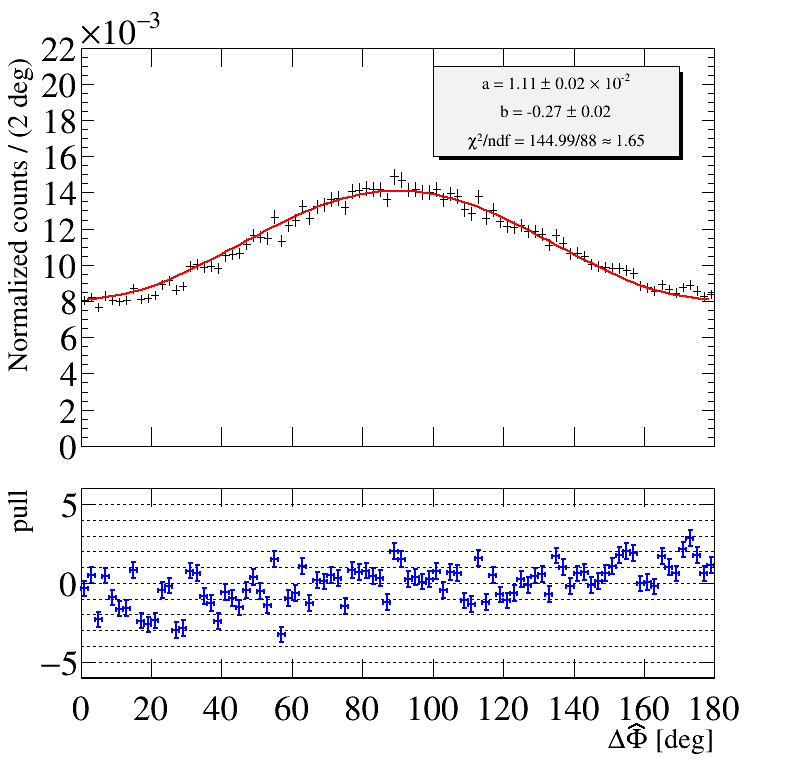}{Experimental data}{fig:correction-r1-fits:experiment}
{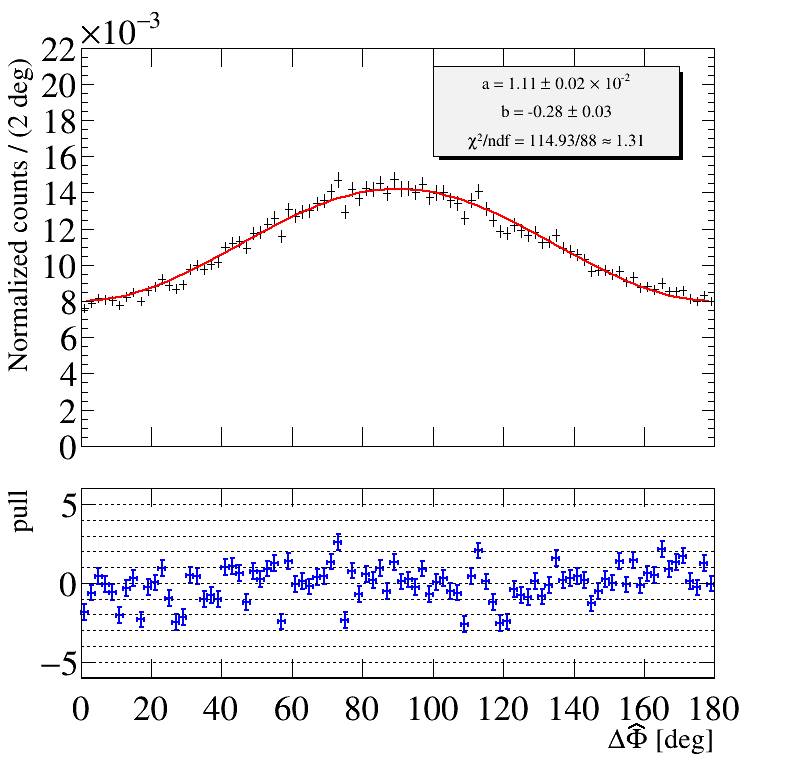}{MC simulations}{fig:correction-r1-fits:simulation}
{The comparison of experimental and MC $\DeltaPhi$ distributions. The spectra are normalized to unity. The simulations are performed for the  $\ket{\psi^{+}}$ polarisation state. Both data and MC samples are pre-filtered using the same selection conditions. The distributions are fitted with the function (red line)   $f(\DeltaPhi) = a(1+b\cos(2\DeltaPhi))$ with the reduced chi-2 equal to $\chi^{2}_{exp} = 144.99/88 \approx 1.65$ and $\chi^{2}_{MC} = 114.93/88 \approx 1.31$ for data and MC respectively. The MC distribution is in good agreement with the data distribution. The fitted values of the $a$ and $b$ parameters are equivalent within the statistical errors.  Pull histogram shows the deviation between the fitted curve and histogram values expressed as $pull(x_i) = \frac{H(x_i)-f(x_i)}{\sigma(x_i)}$, where H is the histogram, the $f$ is the fitted function and $\sigma$ is an error associated with the given bin. }
{fig:correction-r1-fits}{0.5}

Those results can be contrasted with the theoretical value of the maximum visibility squared $\Vis^{2}_{max}=\Vis^{2}(\tTheta_{max}=81.66^\deg)\\ \approx (0.69)^2 \approx 0.48$. However, as explained the analysed sample consists of the events with the range of scattering angles restricted to the circular region fulfilling the condition \refEq{eq:circle-theta-cut}, hence the effective possible value that can be extracted from the experimental data is expected to be smaller than $\Vis^{2}_{max}$. We quantify the upper boundary by determining the weighted mean $\langle\Vis^{2}\rangle_{circle} \approx 0.4$ calculated over the selection region, taking into account the event density extracted from the experimental distribution (see more details in ~\ref{appendix:weighted_mean}).     $\langle \Vis^{2}\rangle_{circle}$ is dominated by the maximum event density point visible in ~\refFig{fig:experimetaldata:tvt:measurements}.

The higher value of $\langle\Vis^{2}\rangle_{circle}$ with respect to the $\Vis^2_{exp}$ could be understood in terms of the detector dilution effects that are not fully compensated by the application of the multiplicative correction coefficients. In principle, the extracted visibility can also be lowered by the presence of a fraction of unpolarized events, or finally, it can vary depending on the initial bipartite polarization state. Those investigations are beyond the scope of the current article.
Here, we would like to emphasise the good agreement between the $\Vis^2_{exp}$ and $\Vis^2_{MC}$.

To further compare the MC to the experimental data, we performed studies on the influence of the high visibility region choice defined in \refEq{eq:circle-theta-cut} by selecting various radii and circle centers. For each region, we calculated the corresponding $\DeltaPhi$ distribution and extracted the fitted squared visibility value together with the corresponding confidence intervals at 95 \% confidence level, as shown in ~\refFig{fig:diff-circle-position}. We also determined the upper boundary $\langle \Vis^{2}\rangle_{circle}$ denoted as a green line.
\addFourFigs
{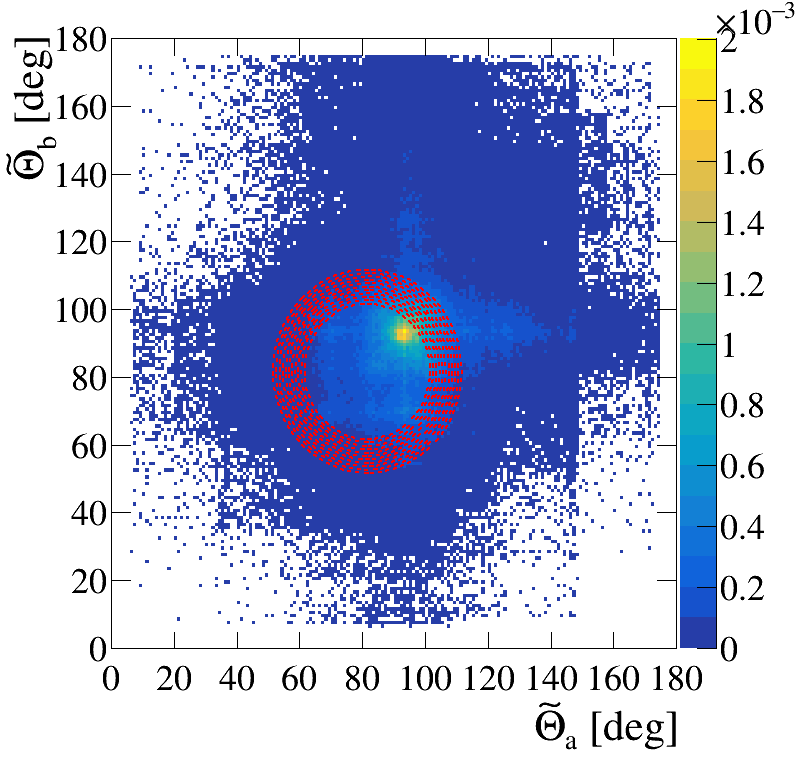}{}{fig:diff-radius-impact:tvt}
{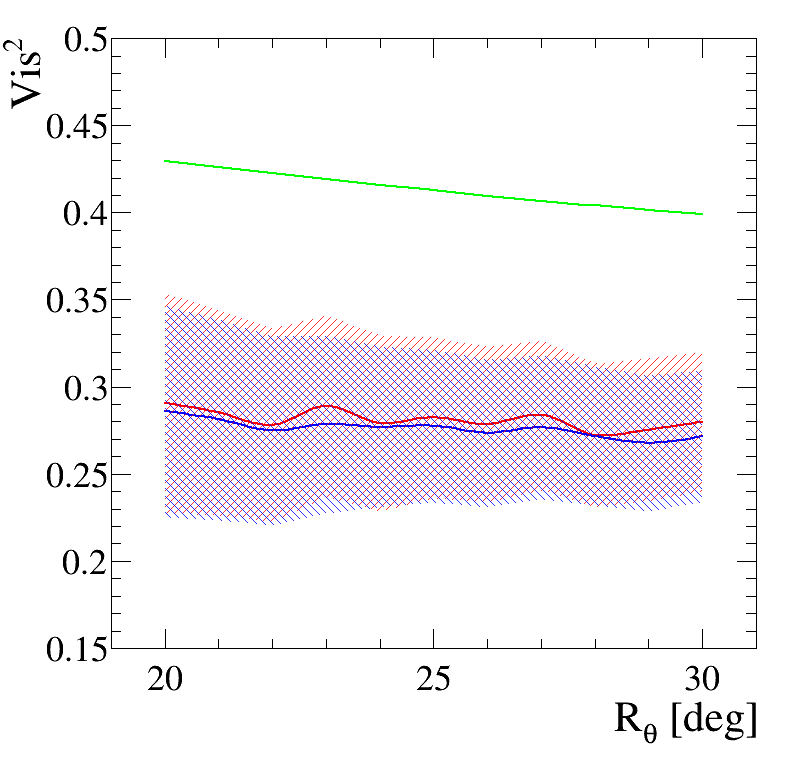}{}{fig:diff-radius-impact:vis2}
{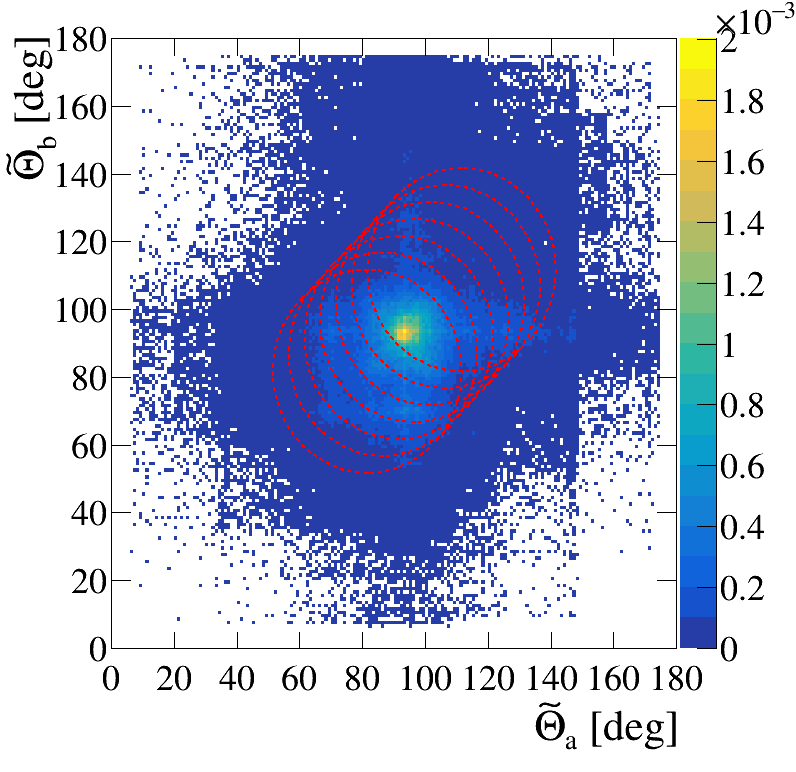}{}{fig:diff-circle-position:tvt}
{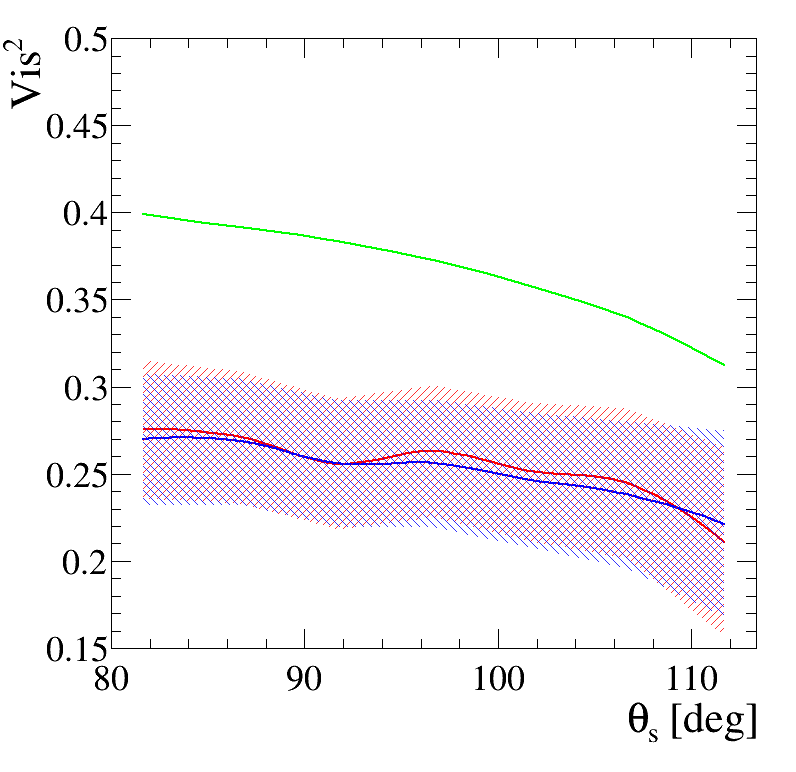}{}{fig:diff-circle-position:vis2}
{The dependency of the $\Vis^2$ determined from the fit of the $\DeltaPhi$ distribution for various choices of the selected visibility regions. 
Figure (a) shows the selection of circular areas with different radii $R_{\tilde{\Theta}_a,\tilde{\Theta}_b} \in [{20^\circ, \dots, 30^\circ}]$ and (c) different circle's centre positions, $\tilde{\Theta}_a=\tilde{\Theta}_b=\theta_s$, with the fixed radius $R_{\tilde{\Theta}_a,\tilde{\Theta}_b}=30^\circ$. 
In Figures (b) and (d), the red and blue lines represent the visibility $\Vis^2$ extracted from a fit to the VW simulation and the experimental data distributions, respectively. The shaded areas correspond to the confidence intervals at a 95\% confidence level. The green line shows the weighted mean  $\langle\Vis^2\rangle_{circle}$ calculated for the given selection region. }
{fig:diff-circle-position}{0.5}
The changes  of the $\Vis^2_{exp}$ and  $\Vis^2_{MC}$ values follow the same trend.  As shown in \refFig{fig:diff-radius-impact:vis2}, lowering the radius, and hence restricting the sample to the scattering angle range closer to the maximum visibility point, increases the extracted visibility. However, the effect is rather mild because of the domination of the highest event density point, which is included in all selected regions. 
The more pronounced change can be observed in \refFig{fig:diff-circle-position:vis2}. Changing the centre of the selection circular region from the $\tTheta_{max}=81.66^\deg $ to higher angular values leads to a decrease of the extracted visibility as predicted by the theory.

\subsection{Background contributions and additional cross-checks}

The main source of background consists of (I) accidental coincidences in which at least one of the four hits comes from different positronium decay,  (II) events with the wrongly associated primary and secondary photon pairs, and (III) events in which one of the hits correspond to the interaction with the deexcitation photon misidentified with the annihilation one. The sample contamination from background events is reduced by applying the aforementioned filtering procedure, e.g. the restriction of the primary hit time difference allows for a great reduction of the background originating from the accidental coincidences.
It is worth pointing out that since the background pairs by definition are not quantum-correlated, the remaining background events will contribute to the $\DeltaPhi$ distribution solely as a uniformly distributed background lowering the overall visibility, but without deforming the $\DeltaPhi$ distributions.

Additional cross-checks were performed to validate the stability of the obtained results.  
The fit to the experimental data is slightly biased as seen in the pull distribution in ~\refFig{fig:correction-r1-fits}, bottom panel. As shown in ~\ref{appendix:fit_systematics}, the presence of the bias has a negligible influence on the results, below the statistical uncertainty.     
The largest correlation between the fit parameters is of the order of 1\% among all tested cases.
The impact of the binning was evaluated by doubling the bin width from 2 to 4 $^{\deg}$. The deviation of the $b$ value is equal to 0.01.

\section{Summary and discussion}
\label{sec:summary}

In this article, we studied the correlations in the polarization degree of freedom for two-photon systems beyond the optical energy regime probed by the double Compton scatterings. Such a system can result from the decay of the para-positronium,  from the direct positron-electron annihilation process or from the ortho-positronium decay via spin-exchange or pick-off processes \refPub{bass_colloquium_nodate}. 

We present the theory of how the experimental observable $\Delta\hat\Phi$, which corresponds to the angle between the two scattering planes, relates to the initial state of the two gammas exploiting the Klein-Nishina formula for double Compton events introduced in Ref.\refPub{quantum1}. 

We introduced the Monte Carlo simulation model, named the Vienna-Warsaw (VW) Simulation Model,  implemented in the Geant4 simulation framework. It can simulate the expected $\Delta\hat\Phi$--distribution for arbitrary initial polarization states. 

We validate the VW Simulation Model by comparing its prediction to the experimental spectra based on data samples measured with the J-PET Big Barrel detector. A good agreement between VW Simulation Model and experimental data is found, as e.g. presented in \refFig{fig:correction-r1-fits}. 
The obtained spectra can be contrasted to the other experimental results performed in recent years~\cite{watts_photon_2021, ivashkinTestingEntanglementAnnihilation2023, parashariClosingDoorPuzzle2024,bordesFirstDetailedStudy2024, moskalNonmaximalEntanglementPhotons2025}. The obtained $\Vis^{2}$ values are presented in~\refFig{fig:comparison}.
As noted in~\cite{bordesFirstDetailedStudy2024}, the overall 
normalisations applied in different setups might be influenced by various effects, including detector-specific conditions. Hence, the absolute comparison should be taken with caution.
\addOneFig{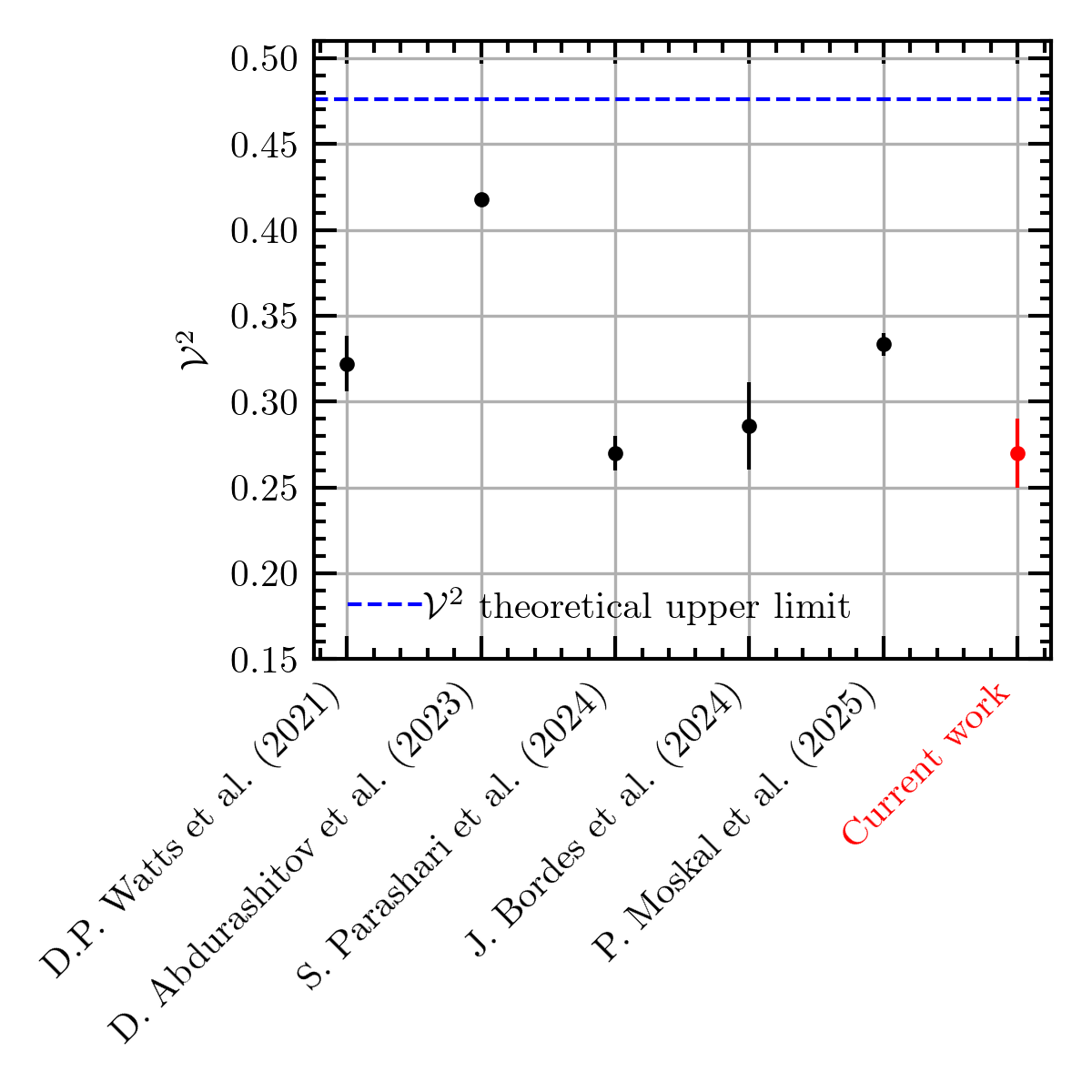}
{Comparison of the experimentally determined squared visibilities $\Vis^{2}$ from recent measurements in the two-photon systems. The comparison must be taken cautiously because the different setup-specific effects can influence the overall normalization~\refPub{bordesFirstDetailedStudy2024}.}
{fig:comparison}{0.5}

We also extracted the values of the squared visibility $\mathcal{V}^2$ from the data for different Compton angle regions (see \refFig{fig:diff-circle-position}). These visibilities show very good agreement between the VW simulation model and the experimental data, and they follow the theoretical prediction. However, the observed visibilities are lower than the estimated theoretical upper bounds, which may be attributed to several factors.
For instance, as discussed in~\refPub{bordesFirstDetailedStudy2024}, accounting for the background caused by multiple scattering - photons undergoing several interactions within the detector crystals -  can lead to an enhancement of the visibility factor.  
An alternative explanation was proposed in~\cite{moskalNonmaximalEntanglementPhotons2025},  where the reduced $\Vis^{2}$ values were linked to the relative probabilities of different positronium annihilation modes in the porous source used in the experiment. Specifically, the authors suggested that photons from pick-off annihilation are not entangled, while those from direct and para-positronium annihilations are maximally entangled. This would naturally result in a lower overall visibility if a significant fraction of events came from non-entangled pick-off annihilation. 
However, in the current study, based on the same experimental input dataset as~\cite{moskalNonmaximalEntanglementPhotons2025}, we demonstrate excellent agreement between the measured data and MC simulations, using a signal model that assumes a maximally-entangled Bell state. This model includes accidental and scattering coincidences as background, but does not account for other positronium decay modes. Thus, our results do not support the hypothesis that the degree of entanglement is significantly affected by the surrounding environment, although they do not categorically rule it out either.
On the other hand, since our Geant4-based simulations do include the contribution from multiple scattering background within the scintillators, it is plausible that properly deconvolving this effect could further enhance the observed visibility factors as previously proposed by~\refPub{bordesFirstDetailedStudy2024}.

In conclusion, those results show that the considered theory is capable of capturing the main underlying physics and that the VW simulation model incorporates the experimental effects efficiently, allowing us to deduce from the experimental data the underlying phenomena.

In the future, we will use this as a starting point to address the foundational problems behind Compton scattering events~\cite{hiesmayrQuantumErrorChannels2024,caradonnaStokesparameterRepresentationCompton2024,tkachevMeasuringEvolutionEntanglement2025} and the physics behind the correlations between the two gamma events and three gamma events~\refPub{hiesmayr_genuine_2017} resulting from positronium. 

\section*{Acknowledgments}
The authors thank E. Czerwiński for his editorial remarks and suggestions. The authors thank P. Moskal and other J-PET collaboration members for providing the input data sample used in the analysis. 
B.C.H. acknowledges gratefully that this research was funded in whole, or in part, by the Austrian Science Fund (FWF) project P36102-N (Grant DOI:10.55776/P36102). 
W.K. acknowledges the support by the Foundation for Polish Science through the FIRST TEAM FENG.02.02-IP.05-0152/23 programme co-financed by the European Union under the European Funds for Smart Economy 2021-2027 (FENG).
This work was completed with resources provided by the Świerk Computing Centre at the National Centre for Nuclear Research.
For the purpose of open access, the author has applied a CC BY public copyright licence to any Author Accepted Manuscript version arising from this submission.
\newpage 
\bibliographystyle{unsrtnat}
\bibliography{final_version}

\begin{thebibliography}{52}
\providecommand{\natexlab}[1]{#1}
\providecommand{\url}[1]{\texttt{#1}}
\expandafter\ifx\csname urlstyle\endcsname\relax
  \providecommand{\doi}[1]{doi: #1}\else
  \providecommand{\doi}{doi: \begingroup \urlstyle{rm}\Url}\fi

\bibitem[Humblet(1943)]{humblet_sur_1943}
J.~Humblet.
\newblock Sur le moment d'impulsion d'une onde électromagnétique.
\newblock \emph{Physica}, 10\penalty0 (7):\penalty0 585--603, July 1943.
\newblock ISSN 00318914.
\newblock \doi{10.1016/S0031-8914(43)90626-3}.
\newblock URL
  \url{https://linkinghub.elsevier.com/retrieve/pii/S0031891443906263}.
\newblock Number: 7.

\bibitem[Akhiezer and Berestetskij(1965)]{singlephoton1}
A.I. Akhiezer and V.B. Berestetskij.
\newblock \emph{Quantum {Electrodynamics}}.
\newblock Interscience Publishers, New York, 1965.

\bibitem[Van~Enk and Nienhuis(1994)]{singlephoton2}
S.J. Van~Enk and G.~Nienhuis.
\newblock Commutation {Rules} and {Eigenvalues} of {Spin} and {Orbital}
  {Angular} {Momentum} of {Radiation} {Fields}.
\newblock \emph{Journal of Modern Optics}, 41\penalty0 (5):\penalty0 963--977,
  May 1994.
\newblock ISSN 0950-0340, 1362-3044.
\newblock \doi{10.1080/09500349414550911}.
\newblock URL
  \url{http://www.tandfonline.com/doi/abs/10.1080/09500349414550911}.

\bibitem[Enk and Nienhuis(1994)]{singlephoton3}
S.~J.~van Enk and G~Nienhuis.
\newblock Spin and {Orbital} {Angular} {Momentum} of {Photons}.
\newblock \emph{Europhys. Lett.}, 25\penalty0 (7):\penalty0 497--501, March
  1994.
\newblock ISSN 0295-5075, 1286-4854.
\newblock \doi{10.1209/0295-5075/25/7/004}.
\newblock URL
  \url{https://iopscience.iop.org/article/10.1209/0295-5075/25/7/004}.

\bibitem[Barnett and Allen(1994)]{singlephoton4}
Stephen~M. Barnett and L.~Allen.
\newblock Orbital angular momentum and nonparaxial light beams.
\newblock \emph{Optics Communications}, 110\penalty0 (5-6):\penalty0 670--678,
  September 1994.
\newblock ISSN 00304018.
\newblock \doi{10.1016/0030-4018(94)90269-0}.
\newblock URL
  \url{https://linkinghub.elsevier.com/retrieve/pii/0030401894902690}.

\bibitem[Bliokh et~al.(2017)Bliokh, Bekshaev, and Nori]{singlephoton5}
Konstantin~Y Bliokh, Aleksandr~Y Bekshaev, and Franco Nori.
\newblock Optical momentum and angular momentum in complex media: from the
  {Abraham}–{Minkowski} debate to unusual properties of surface
  plasmon-polaritons.
\newblock \emph{New J. Phys.}, 19\penalty0 (12):\penalty0 123014, December
  2017.
\newblock ISSN 1367-2630.
\newblock \doi{10.1088/1367-2630/aa8913}.
\newblock URL
  \url{https://iopscience.iop.org/article/10.1088/1367-2630/aa8913}.

\bibitem[Motta et~al.(2019)Motta, Guarnieri, Lanz, and Hiesmayr]{HiesmayrTAM}
M~Motta, G~Guarnieri, L~Lanz, and B~Hiesmayr.
\newblock Unified treatment of the total angular momentum of single photons via
  generalized quantum observables.
\newblock \emph{New J. Phys.}, 21\penalty0 (2):\penalty0 023017, February 2019.
\newblock ISSN 1367-2630.
\newblock \doi{10.1088/1367-2630/aaf841}.
\newblock URL
  \url{https://iopscience.iop.org/article/10.1088/1367-2630/aaf841}.

\bibitem[Madsen et~al.(2022)Madsen, Laudenbach, Askarani, Rortais, Vincent,
  Bulmer, Miatto, Neuhaus, Helt, Collins, Lita, Gerrits, Nam, Vaidya, Menotti,
  Dhand, Vernon, Quesada, and Lavoie]{madsen_quantum_2022}
Lars~S. Madsen, Fabian Laudenbach, Mohsen~Falamarzi. Askarani, Fabien Rortais,
  Trevor Vincent, Jacob F.~F. Bulmer, Filippo~M. Miatto, Leonhard Neuhaus,
  Lukas~G. Helt, Matthew~J. Collins, Adriana~E. Lita, Thomas Gerrits, Sae~Woo
  Nam, Varun~D. Vaidya, Matteo Menotti, Ish Dhand, Zachary Vernon, Nicolás
  Quesada, and Jonathan Lavoie.
\newblock Quantum computational advantage with a programmable photonic
  processor.
\newblock \emph{Nature}, 606\penalty0 (7912):\penalty0 75--81, June 2022.
\newblock ISSN 0028-0836, 1476-4687.
\newblock \doi{10.1038/s41586-022-04725-x}.
\newblock URL \url{https://www.nature.com/articles/s41586-022-04725-x}.
\newblock Number: 7912.

\bibitem[Hiesmayr(2022)]{MLHiesmayr}
Beatrix~C. Hiesmayr.
\newblock A {Quantum} {Information} {Theoretic} {View} {On} {A} {Deep}
  {Quantum} {Neural} {Network}.
\newblock \emph{AIP Conference Proceedings}, 2022.
\newblock \doi{10.48550/ARXIV.2212.12906}.
\newblock URL \url{https://arxiv.org/abs/2212.12906}.
\newblock Publisher: arXiv Version Number: 2.

\bibitem[Biamonte et~al.(2017)Biamonte, Wittek, Pancotti, Rebentrost, Wiebe,
  and Lloyd]{biamonte_quantum_2017}
Jacob Biamonte, Peter Wittek, Nicola Pancotti, Patrick Rebentrost, Nathan
  Wiebe, and Seth Lloyd.
\newblock Quantum machine learning.
\newblock \emph{Nature}, 549\penalty0 (7671):\penalty0 195--202, September
  2017.
\newblock ISSN 0028-0836, 1476-4687.
\newblock \doi{10.1038/nature23474}.
\newblock URL \url{http://www.nature.com/articles/nature23474}.
\newblock Number: 7671.

\bibitem[Wan et~al.(2017)Wan, Dahlsten, Kristjánsson, Gardner, and
  Kim]{wan_quantum_2017}
Kwok~Ho Wan, Oscar Dahlsten, Hlér Kristjánsson, Robert Gardner, and M.~S.
  Kim.
\newblock Quantum generalisation of feedforward neural networks.
\newblock \emph{npj Quantum Inf}, 3\penalty0 (1):\penalty0 36, September 2017.
\newblock ISSN 2056-6387.
\newblock \doi{10.1038/s41534-017-0032-4}.
\newblock URL \url{https://www.nature.com/articles/s41534-017-0032-4}.
\newblock Number: 1.

\bibitem[Bennett and Brassard(2014)]{BB84}
Charles~H. Bennett and Gilles Brassard.
\newblock Quantum cryptography: {Public} key distribution and coin tossing.
\newblock \emph{Theoretical Computer Science}, 560:\penalty0 7--11, December
  2014.
\newblock ISSN 03043975.
\newblock \doi{10.1016/j.tcs.2014.05.025}.
\newblock URL
  \url{https://linkinghub.elsevier.com/retrieve/pii/S0304397514004241}.

\bibitem[Ekert(1991)]{E91}
Artur~K. Ekert.
\newblock Quantum cryptography based on {Bell}’s theorem.
\newblock \emph{Phys. Rev. Lett.}, 67\penalty0 (6):\penalty0 661--663, August
  1991.
\newblock ISSN 0031-9007.
\newblock \doi{10.1103/PhysRevLett.67.661}.
\newblock URL \url{https://link.aps.org/doi/10.1103/PhysRevLett.67.661}.

\bibitem[Schauer et~al.(2010)Schauer, Huber, and Hiesmayr]{SecretSharing}
Stefan Schauer, Marcus Huber, and Beatrix~C. Hiesmayr.
\newblock Experimentally feasible security check for n -qubit quantum secret
  sharing.
\newblock \emph{Phys. Rev. A}, 82\penalty0 (6):\penalty0 062311, December 2010.
\newblock ISSN 1050-2947, 1094-1622.
\newblock \doi{10.1103/PhysRevA.82.062311}.
\newblock URL \url{https://link.aps.org/doi/10.1103/PhysRevA.82.062311}.

\bibitem[Bennett et~al.(1992)Bennett, Bessette, Brassard, Salvail, and
  Smolin]{ExpQC}
Charles~H. Bennett, François Bessette, Gilles Brassard, Louis Salvail, and
  John Smolin.
\newblock Experimental quantum cryptography.
\newblock \emph{J. Cryptology}, 5\penalty0 (1):\penalty0 3--28, January 1992.
\newblock ISSN 0933-2790, 1432-1378.
\newblock \doi{10.1007/BF00191318}.
\newblock URL \url{http://link.springer.com/10.1007/BF00191318}.

\bibitem[Cao et~al.(2020)Cao, Li, Yang, Jiang, Li, Hu, Abulizi, Li, Zhang, Sun,
  Liu, Jiang, Liao, Ren, Li, You, Wang, Yin, Lu, Wang, Zhang, Peng, and
  Pan]{LongDistance}
Yuan Cao, Yu-Huai Li, Kui-Xing Yang, Yang-Fan Jiang, Shuang-Lin Li, Xiao-Long
  Hu, Maimaiti Abulizi, Cheng-Long Li, Weijun Zhang, Qi-Chao Sun, Wei-Yue Liu,
  Xiao Jiang, Sheng-Kai Liao, Ji-Gang Ren, Hao Li, Lixing You, Zhen Wang, Juan
  Yin, Chao-Yang Lu, Xiang-Bin Wang, Qiang Zhang, Cheng-Zhi Peng, and Jian-Wei
  Pan.
\newblock Long-{Distance} {Free}-{Space} {Measurement}-{Device}-{Independent}
  {Quantum} {Key} {Distribution}.
\newblock \emph{Phys. Rev. Lett.}, 125\penalty0 (26):\penalty0 260503, December
  2020.
\newblock ISSN 0031-9007, 1079-7114.
\newblock \doi{10.1103/PhysRevLett.125.260503}.
\newblock URL \url{https://link.aps.org/doi/10.1103/PhysRevLett.125.260503}.

\bibitem[McNamara et~al.(2014)McNamara, Toghyani, Gillam, Wu, and
  Kuncic]{mcnamara_towards_2014}
A~L McNamara, M~Toghyani, J~E Gillam, K~Wu, and Z~Kuncic.
\newblock Towards optimal imaging with {PET}: an \textit{in silico} feasibility
  study.
\newblock \emph{Phys. Med. Biol.}, 59\penalty0 (24):\penalty0 7587--7600,
  December 2014.
\newblock ISSN 0031-9155, 1361-6560.
\newblock \doi{10.1088/0031-9155/59/24/7587}.
\newblock URL
  \url{https://iopscience.iop.org/article/10.1088/0031-9155/59/24/7587}.
\newblock Number: 24.

\bibitem[Toghyani et~al.(2016)Toghyani, Gillam, McNamara, and
  Kuncic]{toghyani_polarisation-based_2016}
M~Toghyani, J~E Gillam, A~L McNamara, and Z~Kuncic.
\newblock Polarisation-based coincidence event discrimination: an \textit{in
  silico} study towards a feasible scheme for {Compton}-{PET}.
\newblock \emph{Phys. Med. Biol.}, 61\penalty0 (15):\penalty0 5803--5817,
  August 2016.
\newblock ISSN 0031-9155, 1361-6560.
\newblock \doi{10.1088/0031-9155/61/15/5803}.
\newblock URL
  \url{https://iopscience.iop.org/article/10.1088/0031-9155/61/15/5803}.
\newblock Number: 15.

\bibitem[Moskal et~al.(2018)Moskal, Krawczyk, Hiesmayr, Bała, Curceanu,
  Czerwiński, Dulski, Gajos, Gorgol, Del~Grande, Jasińska, Kacprzak, Kapłon,
  Kisielewska, Klimaszewski, Korcyl, Kowalski, Kozik, Krzemień, Kubicz,
  Mohammed, Niedźwiecki, Pałka, Pawlik-Niedźwiecka, Raczyński, Raj, Rudy,
  Sharma, Silarski, {Shivani}, Shopa, Skurzok, Wiślicki, and
  Zgardzińska]{quantum2}
P.~Moskal, N.~Krawczyk, B.~C. Hiesmayr, M.~Bała, C.~Curceanu, E.~Czerwiński,
  K.~Dulski, A.~Gajos, M.~Gorgol, R.~Del~Grande, B.~Jasińska, K.~Kacprzak,
  L.~Kapłon, D.~Kisielewska, K.~Klimaszewski, G.~Korcyl, P.~Kowalski,
  T.~Kozik, W.~Krzemień, E.~Kubicz, M.~Mohammed, Sz. Niedźwiecki, M.~Pałka,
  M.~Pawlik-Niedźwiecka, L.~Raczyński, J.~Raj, Z.~Rudy, S.~Sharma,
  M.~Silarski, {Shivani}, R.~Y. Shopa, M.~Skurzok, W.~Wiślicki, and
  B.~Zgardzińska.
\newblock Feasibility studies of the polarization of photons beyond the optical
  wavelength regime with the {J}-{PET} detector.
\newblock \emph{Eur. Phys. J. C}, 78\penalty0 (11):\penalty0 970, November
  2018.
\newblock ISSN 1434-6044, 1434-6052.
\newblock \doi{10.1140/epjc/s10052-018-6461-1}.
\newblock URL \url{http://link.springer.com/10.1140/epjc/s10052-018-6461-1}.

\bibitem[Bass et~al.()Bass, Moskal, Mariazzi, and
  Stepien]{bass_colloquium_nodate}
Steven~D. Bass, Pawel Moskal, Sebastiano Mariazzi, and Ewa Stepien.
\newblock Colloquium: {Positronium} physics and biomedical applications.
\newblock URL \url{https://arxiv.org/abs/2302.09246}.

\bibitem[Watts(2024)]{wattsQuantumEntangledPET2024}
D.~P. Watts.
\newblock Quantum entangled pet - recent progress and new insights.
\newblock In \emph{2024 IEEE Nuclear Science Symposium (NSS), Medical Imaging
  Conference (MIC) and Room Temperature Semiconductor Detector Conference
  (RTSD)}, pages 1--1, Tampa, FL, USA, October 2024. IEEE.
\newblock ISBN 9798350388152.
\newblock \doi{10.1109/NSS/MIC/RTSD57108.2024.10658587}.

\bibitem[Kožuljević et~al.(2021)Kožuljević, Bosnar, Kuncic, Makek,
  Parashari, and Žugec]{kozuljevic_study_2021}
Ana~Marija Kožuljević, Damir Bosnar, Zdenka Kuncic, Mihael Makek, Siddharth
  Parashari, and Petar Žugec.
\newblock Study of {Multi}-{Pixel} {Scintillator} {Detector} {Configurations}
  for {Measuring} {Polarized} {Gamma} {Radiation}.
\newblock \emph{Condensed Matter}, 6\penalty0 (4):\penalty0 43, November 2021.
\newblock ISSN 2410-3896.
\newblock \doi{10.3390/condmat6040043}.
\newblock URL \url{https://www.mdpi.com/2410-3896/6/4/43}.
\newblock Number: 4.

\bibitem[Makek et~al.(2020)Makek, Bosnar, Pavelić, Šenjug, and
  Žugec]{makek_single-layer_2020}
Mihael Makek, Damir Bosnar, Luka Pavelić, Pavla Šenjug, and Petar Žugec.
\newblock Single-layer {Compton} detectors for measurement of polarization
  correlations of annihilation quanta.
\newblock \emph{Nuclear Instruments and Methods in Physics Research Section A:
  Accelerators, Spectrometers, Detectors and Associated Equipment},
  958:\penalty0 162835, April 2020.
\newblock ISSN 01689002.
\newblock \doi{10.1016/j.nima.2019.162835}.
\newblock URL
  \url{https://linkinghub.elsevier.com/retrieve/pii/S0168900219312707}.

\bibitem[Watts et~al.(2021)Watts, Bordes, Brown, Cherlin, Newton, Allison,
  Bashkanov, Efthimiou, and Zachariou]{watts_photon_2021}
D.~P. Watts, J.~Bordes, J.~R. Brown, A.~Cherlin, R.~Newton, J.~Allison,
  M.~Bashkanov, N.~Efthimiou, and N.~A. Zachariou.
\newblock Photon quantum entanglement in the {MeV} regime and its application
  in {PET} imaging.
\newblock \emph{Nat Commun}, 12\penalty0 (1):\penalty0 2646, May 2021.
\newblock ISSN 2041-1723.
\newblock \doi{10.1038/s41467-021-22907-5}.
\newblock URL \url{https://www.nature.com/articles/s41467-021-22907-5}.
\newblock Number: 1.

\bibitem[Abdurashitov et~al.(2022)Abdurashitov, Baranov, Borisenko, Guber,
  Ivashkin, Morozov, Musin, Strizhak, Tkachev, Volkov, and
  Zhuikov]{abdurashitov_setup_2022}
D.~Abdurashitov, A.~Baranov, D.~Borisenko, F.~Guber, A.~Ivashkin, S.~Morozov,
  S.~Musin, A.~Strizhak, I.~Tkachev, V.~Volkov, and B.~Zhuikov.
\newblock Setup of {Compton} polarimeters for measuring entangled annihilation
  photons.
\newblock \emph{J. Inst.}, 17\penalty0 (03):\penalty0 P03010, March 2022.
\newblock ISSN 1748-0221.
\newblock \doi{10.1088/1748-0221/17/03/P03010}.
\newblock URL
  \url{https://iopscience.iop.org/article/10.1088/1748-0221/17/03/P03010}.
\newblock Number: 03.

\bibitem[Ivashkin et~al.(2023)Ivashkin, Abdurashitov, Baranov, Guber, Morozov,
  Musin, Strizhak, and chev]{ivashkinTestingEntanglementAnnihilation2023}
Alexander Ivashkin, Dzhonrid Abdurashitov, Alexander Baranov, Fedor Guber,
  Sergey Morozov, Sultan Musin, Alexander Strizhak, and Igor chev.
\newblock Testing entanglement of annihilation photons.
\newblock \emph{Sci Rep}, 13\penalty0 (1):\penalty0 7559, May 2023.
\newblock ISSN 2045-2322.
\newblock \doi{10.1038/s41598-023-34767-8}.

\bibitem[Parashari et~al.(2024)Parashari, Bosnar, Fri{\v s}{\v c}i{\'c}, Ko{\v
  z}uljevi{\'c}, Kuncic, {\v Z}ugec, and Makek]{parashariClosingDoorPuzzle2024}
Siddharth Parashari, Damir Bosnar, Ivica Fri{\v s}{\v c}i{\'c}, Ana~Marija
  Ko{\v z}uljevi{\'c}, Zdenka Kuncic, Petar {\v Z}ugec, and Mihael Makek.
\newblock Closing the door on the ``puzzle of decoherence'' of annihilation
  quanta.
\newblock \emph{Physics Letters B}, 852:\penalty0 138628, May 2024.
\newblock ISSN 03702693.
\newblock \doi{10.1016/j.physletb.2024.138628}.

\bibitem[Bordes et~al.(2024)Bordes, Brown, Watts, Bashkanov, Gibson, Newton,
  and Zachariou]{bordesFirstDetailedStudy2024}
Julien Bordes, James~R. Brown, Daniel~P. Watts, Mikhail Bashkanov, Kieran
  Gibson, Ruth Newton, and Nicholas Zachariou.
\newblock First detailed study of the quantum decoherence of entangled gamma
  photons.
\newblock \emph{Phys. Rev. Lett.}, 133\penalty0 (13):\penalty0 132502,
  September 2024.
\newblock ISSN 0031-9007, 1079-7114.
\newblock \doi{10.1103/PhysRevLett.133.132502}.

\bibitem[Kim et~al.(2023)Kim, Rachman, Taisei, Uenomachi, Shimazoe, and
  Takahashi]{kimBackgroundReductionPET2023}
Donghwan Kim, Agus~Nur Rachman, Ueki Taisei, Mizuki Uenomachi, Kenji Shimazoe,
  and Hiroyuki Takahashi.
\newblock Background reduction in pet by double compton scattering of quantum
  entangled annihilation photons.
\newblock \emph{J. Inst.}, 18\penalty0 (07):\penalty0 P07007, July 2023.
\newblock ISSN 1748-0221.
\newblock \doi{10.1088/1748-0221/18/07/P07007}.

\bibitem[Shoop et~al.(2024)Shoop, Romanchek, Gholami, Kuo, Kupinski, King,
  Furenlid, and Abbaszadeh]{shoopPETSimulationStudy2024}
G.~Shoop, G.~Romanchek, K.~Gholami, P.~H. Kuo, M.~Kupinski, M.~A. King, L.~R.
  Furenlid, and S.~Abbaszadeh.
\newblock Pet simulation study of quantum entangled annihilation photons for
  application in gamma-positron imaging.
\newblock In \emph{2024 IEEE Nuclear Science Symposium (NSS), Medical Imaging
  Conference (MIC) and Room Temperature Semiconductor Detector Conference
  (RTSD)}, pages 1--2, Tampa, FL, USA, October 2024. IEEE.
\newblock ISBN 9798350388152.
\newblock \doi{10.1109/NSS/MIC/RTSD57108.2024.10657853}.

\bibitem[Das et~al.(2024)Das, Verma, and Prasad]{dasExploringPETImaging2024}
Pragya Das, Ritesh Verma, and Kaustav Prasad.
\newblock Exploring pet imaging with scattered photons and polarization
  characteristics.
\newblock \emph{Bio-Algorithms and Med-Systems}, 20\penalty0 (Special
  Issue):\penalty0 10--16, November 2024.
\newblock ISSN 1896-530X, 1895-9091.
\newblock \doi{10.5604/01.3001.0054.8576}.

\bibitem[Ko{\v z}uljevi{\'c} et~al.(2024)Ko{\v z}uljevi{\'c}, Bokuli{\'c},
  Gro{\v s}ev, Kuncic, Parashari, Paveli{\'c}, and
  Makek]{kozuljevicInvestigationSpatialResolution2024}
Ana~Marija Ko{\v z}uljevi{\'c}, Tomislav Bokuli{\'c}, Darko Gro{\v s}ev, Zdenka
  Kuncic, Siddharth Parashari, Luka Paveli{\'c}, and Mihael Makek.
\newblock Investigation of the spatial resolution of pet imaging system
  measuring polarization-correlated compton events.
\newblock \emph{Nuclear Instruments and Methods in Physics Research Section A:
  Accelerators, Spectrometers, Detectors and Associated Equipment},
  1068:\penalty0 169795, November 2024.
\newblock ISSN 01689002.
\newblock \doi{10.1016/j.nima.2024.169795}.

\bibitem[Moulin et~al.(2024)Moulin, Mont{\'e}mont, Stanchina, and
  L{\'e}tang]{moulinBenefitsQuantumEntanglement2024}
M.~Moulin, G.~Mont{\'e}mont, S.~Stanchina, and J.M. L{\'e}tang.
\newblock On the benefits of quantum entanglement information for semiconductor
  pet imaging.
\newblock In \emph{2024 IEEE Nuclear Science Symposium (NSS), Medical Imaging
  Conference (MIC) and Room Temperature Semiconductor Detector Conference
  (RTSD)}, pages 1--1, Tampa, FL, USA, October 2024. IEEE.
\newblock ISBN 9798350388152.
\newblock \doi{10.1109/NSS/MIC/RTSD57108.2024.10657259}.

\bibitem[Moskal et~al.(2025)Moskal, Kumar, Sharma, Beyene, Chug, Coussat,
  Curceanu, Czerwinski, Das, Dulski, Gorgol, Jasinska, Kacprzak, Kaplanoglu,
  Kaplon, Kozik, Lisowski, Lisowski, Mryka, Niedzwiecki, Parzych, del Rio,
  Radler, Skurzok, Stepien, Tanty, Ardebili, and
  Eliyan]{moskalNonmaximalEntanglementPhotons2025}
P.~Moskal, D.~Kumar, S.~Sharma, E.~Y. Beyene, N.~Chug, A.~Coussat, C.~Curceanu,
  E.~Czerwinski, M.~Das, K.~Dulski, M.~Gorgol, B.~Jasinska, K.~Kacprzak,
  T.~Kaplanoglu, L.~Kaplon, T.~Kozik, E.~Lisowski, F.~Lisowski, W.~Mryka,
  S.~Niedzwiecki, S.~Parzych, E.~P. del Rio, M.~Radler, M.~Skurzok, E.~L.
  Stepien, P.~Tanty, K.~Tayefi Ardebili, and K.~Valsan Eliyan.
\newblock Non-maximal entanglement of photons from positron-electron
  annihilation demonstrated using a novel plastic pet scanner, 2025.
\newblock URL \url{https://arxiv.org/abs/2407.08574}.

\bibitem[Hiesmayr and Moskal(2019)]{quantum1}
Beatrix~C. Hiesmayr and Pawel Moskal.
\newblock Witnessing {Entanglement} {In} {Compton} {Scattering} {Processes}
  {Via} {Mutually} {Unbiased} {Bases}.
\newblock \emph{Sci Rep}, 9\penalty0 (1):\penalty0 8166, June 2019.
\newblock ISSN 2045-2322.
\newblock \doi{10.1038/s41598-019-44570-z}.
\newblock URL \url{https://www.nature.com/articles/s41598-019-44570-z}.

\bibitem[Niedźwiecki et~al.(2017)Niedźwiecki, Białas, Curceanu, Czerwiński,
  Dulski, Gajos, Głowacz, Gorgol, Hiesmayr, Jasińska, Kapłon,
  Kisielewska-Kamińska, Korcyl, Kowalski, Kozik, Krawczyk, Krzemień, Kubicz,
  Mohammed, Pawlik-Niedźwiecka, Pałka, Raczyński, Rudy, Sharma, Sharma,
  Shopa, Silarski, Skurzok, Wieczorek, Wiślicki, Zgardzińska, Zieliński, and
  Moskal]{niedzwiecki_j-pet_2017}
S.~Niedźwiecki, P.~Białas, C.~Curceanu, E.~Czerwiński, K.~Dulski, A.~Gajos,
  B.~Głowacz, M.~Gorgol, B.C. Hiesmayr, B.~Jasińska, {\L{}}.~Kapłon,
  D.~Kisielewska-Kamińska, G.~Korcyl, P.~Kowalski, T.~Kozik, N.~Krawczyk,
  W.~Krzemień, E.~Kubicz, M.~Mohammed, M.~Pawlik-Niedźwiecka, M.~Pałka,
  L.~Raczyński, Z.~Rudy, N.G. Sharma, S.~Sharma, R.Y. Shopa, M.~Silarski,
  M.~Skurzok, A.~Wieczorek, W.~Wiślicki, B.~Zgardzińska, M.~Zieliński, and
  P.~Moskal.
\newblock J-{PET}: {A} {New} {Technology} for the {Whole}-body {PET} {Imaging}.
\newblock \emph{Acta Phys. Pol. B}, 48\penalty0 (10):\penalty0 1567, 2017.
\newblock ISSN 0587-4254, 1509-5770.
\newblock \doi{10.5506/APhysPolB.48.1567}.
\newblock URL
  \url{http://www.actaphys.uj.edu.pl/findarticle?series=Reg&vol=48&page=1567}.

\bibitem[Dulski et~al.(2021)Dulski, Bass, Chhokar, Chug, Curceanu,
  Czerwi{\'n}ski, Dagdar, Gajewski, Gajos, Gorgol, Del~Grande, Hiesmayr,
  Jasi{\'n}ska, Kacprzak, Kap{\l}on, Karimi, Kisielewska, Klimaszewski, Kopka,
  Korcyl, Kowalski, Kozik, Krawczyk, Krzemie{\'n}, Kubicz, Ma{\l}czak,
  Mohammed, Nied{\'z}wiecki, Pa{\l}ka, {Pawlik-Nied{\'z}wiecka}, P{\k
  e}dziwiatr, Raczy{\'n}ski, Raj, Ruci{\'n}ski, Sharma, {Shivani}, Shopa,
  Silarski, Skurzok, St{\k e}pie{\'n}, Tayefi, Wi{\'s}licki, Zgardzi{\'n}ska,
  and Moskal]{dulskiJPETDetectorTool2021a}
K.~Dulski, S.D. Bass, J.~Chhokar, N.~Chug, C.~Curceanu, E.~Czerwi{\'n}ski,
  M.~Dagdar, J.~Gajewski, A.~Gajos, M.~Gorgol, R.~Del~Grande, B.C. Hiesmayr,
  B.~Jasi{\'n}ska, K.~Kacprzak, {\L}.~Kap{\l}on, H.~Karimi, D.~Kisielewska,
  K.~Klimaszewski, P.~Kopka, G.~Korcyl, P.~Kowalski, T.~Kozik, N.~Krawczyk,
  W.~Krzemie{\'n}, E.~Kubicz, P.~Ma{\l}czak, M.~Mohammed, {\relax
  Sz}.~Nied{\'z}wiecki, M.~Pa{\l}ka, M.~{Pawlik-Nied{\'z}wiecka}, M.~P{\k
  e}dziwiatr, L.~Raczy{\'n}ski, J.~Raj, A.~Ruci{\'n}ski, S.~Sharma, {Shivani},
  R.Y. Shopa, M.~Silarski, M.~Skurzok, E.{\L}. St{\k e}pie{\'n}, F.~Tayefi,
  W.~Wi{\'s}licki, B.~Zgardzi{\'n}ska, and P.~Moskal.
\newblock The j-pet detector---a tool for precision studies of
  ortho-positronium decays.
\newblock \emph{Nuclear Instruments and Methods in Physics Research Section A:
  Accelerators, Spectrometers, Detectors and Associated Equipment},
  1008:\penalty0 165452, August 2021.
\newblock ISSN 01689002.
\newblock \doi{10.1016/j.nima.2021.165452}.

\bibitem[Klein and Nishina(1928)]{klein_scattering_1928}
O.~Klein and Y.~Nishina.
\newblock The {Scattering} of {Light} by {Free} {Electrons} according to
  {Dirac}'s {New} {Relativistic} {Dynamics}.
\newblock \emph{Nature}, 122\penalty0 (3072):\penalty0 398--399, September
  1928.
\newblock ISSN 0028-0836, 1476-4687.
\newblock \doi{10.1038/122398b0}.
\newblock URL \url{https://www.nature.com/articles/122398b0}.
\newblock Number: 3072.

\bibitem[Baumgartner et~al.(2007)Baumgartner, Hiesmayr, and
  Narnhofer]{MagicSimplex1}
Bernhard Baumgartner, Beatrix Hiesmayr, and Heide Narnhofer.
\newblock A special simplex in the state space for entangled qudits.
\newblock \emph{J. Phys. A: Math. Theor.}, 40\penalty0 (28):\penalty0
  7919--7938, July 2007.
\newblock ISSN 1751-8113, 1751-8121.
\newblock \doi{10.1088/1751-8113/40/28/S03}.
\newblock URL
  \url{https://iopscience.iop.org/article/10.1088/1751-8113/40/28/S03}.

\bibitem[Hiesmayr(2021)]{MagicSimplex2}
Beatrix~C. Hiesmayr.
\newblock Free versus bound entanglement, a {NP}-hard problem tackled by
  machine learning.
\newblock \emph{Sci Rep}, 11\penalty0 (1):\penalty0 19739, October 2021.
\newblock ISSN 2045-2322.
\newblock \doi{10.1038/s41598-021-98523-6}.
\newblock URL \url{https://www.nature.com/articles/s41598-021-98523-6}.

\bibitem[Allison et~al.(2016)Allison, Amako, Apostolakis, Arce, Asai, Aso,
  Bagli, Bagulya, Banerjee, Barrand, Beck, Bogdanov, Brandt, Brown, Burkhardt,
  Canal, Cano-Ott, Chauvie, Cho, Cirrone, Cooperman, Cortés-Giraldo, Cosmo,
  Cuttone, Depaola, Desorgher, Dong, Dotti, Elvira, Folger, Francis, Galoyan,
  Garnier, Gayer, Genser, Grichine, Guatelli, Guèye, Gumplinger, Howard,
  Hřivnáčová, Hwang, Incerti, Ivanchenko, Ivanchenko, Jones, Jun,
  Kaitaniemi, Karakatsanis, Karamitros, Kelsey, Kimura, Koi, Kurashige,
  Lechner, Lee, Longo, Maire, Mancusi, Mantero, Mendoza, Morgan, Murakami,
  Nikitina, Pandola, Paprocki, Perl, Petrović, Pia, Pokorski, Quesada, Raine,
  Reis, Ribon, Ristić~Fira, Romano, Russo, Santin, Sasaki, Sawkey, Shin,
  Strakovsky, Taborda, Tanaka, Tomé, Toshito, Tran, Truscott, Urban, Uzhinsky,
  Verbeke, Verderi, Wendt, Wenzel, Wright, Wright, Yamashita, Yarba, and
  Yoshida]{allison2016}
J.~Allison, K.~Amako, J.~Apostolakis, P.~Arce, M.~Asai, T.~Aso, E.~Bagli,
  A.~Bagulya, S.~Banerjee, G.~Barrand, B.R. Beck, A.G. Bogdanov, D.~Brandt,
  J.M.C. Brown, H.~Burkhardt, Ph. Canal, D.~Cano-Ott, S.~Chauvie, K.~Cho,
  G.A.P. Cirrone, G.~Cooperman, M.A. Cortés-Giraldo, G.~Cosmo, G.~Cuttone,
  G.~Depaola, L.~Desorgher, X.~Dong, A.~Dotti, V.D. Elvira, G.~Folger,
  Z.~Francis, A.~Galoyan, L.~Garnier, M.~Gayer, K.L. Genser, V.M. Grichine,
  S.~Guatelli, P.~Guèye, P.~Gumplinger, A.S. Howard, I.~Hřivnáčová,
  S.~Hwang, S.~Incerti, A.~Ivanchenko, V.N. Ivanchenko, F.W. Jones, S.Y. Jun,
  P.~Kaitaniemi, N.~Karakatsanis, M.~Karamitros, M.~Kelsey, A.~Kimura, T.~Koi,
  H.~Kurashige, A.~Lechner, S.B. Lee, F.~Longo, M.~Maire, D.~Mancusi,
  A.~Mantero, E.~Mendoza, B.~Morgan, K.~Murakami, T.~Nikitina, L.~Pandola,
  P.~Paprocki, J.~Perl, I.~Petrović, M.G. Pia, W.~Pokorski, J.M. Quesada,
  M.~Raine, M.A. Reis, A.~Ribon, A.~Ristić~Fira, F.~Romano, G.~Russo,
  G.~Santin, T.~Sasaki, D.~Sawkey, J.I. Shin, I.I. Strakovsky, A.~Taborda,
  S.~Tanaka, B.~Tomé, T.~Toshito, H.N. Tran, P.R. Truscott, L.~Urban,
  V.~Uzhinsky, J.M. Verbeke, M.~Verderi, B.L. Wendt, H.~Wenzel, D.H. Wright,
  D.M. Wright, T.~Yamashita, J.~Yarba, and H.~Yoshida.
\newblock Recent developments in {Geant4}.
\newblock \emph{Nuclear Instruments and Methods in Physics Research Section A:
  Accelerators, Spectrometers, Detectors and Associated Equipment},
  835:\penalty0 186--225, November 2016.
\newblock ISSN 01689002.
\newblock \doi{10.1016/j.nima.2016.06.125}.
\newblock URL
  \url{https://linkinghub.elsevier.com/retrieve/pii/S0168900216306957}.

\bibitem[Depaola(2003)]{depaola_new_2003}
G.O. Depaola.
\newblock New {Monte} {Carlo} method for {Compton} and {Rayleigh} scattering by
  polarized gamma rays.
\newblock \emph{Nuclear Instruments and Methods in Physics Research Section A:
  Accelerators, Spectrometers, Detectors and Associated Equipment},
  512\penalty0 (3):\penalty0 619--630, October 2003.
\newblock ISSN 01689002.
\newblock \doi{10.1016/S0168-9002(03)02050-3}.
\newblock URL
  \url{https://linkinghub.elsevier.com/retrieve/pii/S0168900203020503}.

\bibitem[Agostinelli et~al.(2003)Agostinelli, Allison, Amako, Apostolakis,
  Araujo, Arce, Asai, Axen, Banerjee, Barrand, Behner, Bellagamba, Boudreau,
  Broglia, Brunengo, Burkhardt, Chauvie, Chuma, Chytracek, Cooperman, Cosmo,
  Degtyarenko, Dell'Acqua, Depaola, Dietrich, Enami, Feliciello, Ferguson,
  Fesefeldt, Folger, Foppiano, Forti, Garelli, Giani, Giannitrapani, Gibin,
  Gómez~Cadenas, González, Gracia~Abril, Greeniaus, Greiner, Grichine,
  Grossheim, Guatelli, Gumplinger, Hamatsu, Hashimoto, Hasui, Heikkinen,
  Howard, Ivanchenko, Johnson, Jones, Kallenbach, Kanaya, Kawabata, Kawabata,
  Kawaguti, Kelner, Kent, Kimura, Kodama, Kokoulin, Kossov, Kurashige, Lamanna,
  Lampén, Lara, Lefebure, Lei, Liendl, Lockman, Longo, Magni, Maire,
  Medernach, Minamimoto, Mora~de Freitas, Morita, Murakami, Nagamatu, Nartallo,
  Nieminen, Nishimura, Ohtsubo, Okamura, O'Neale, Oohata, Paech, Perl,
  Pfeiffer, Pia, Ranjard, Rybin, Sadilov, Di~Salvo, Santin, Sasaki, Savvas,
  Sawada, Scherer, Sei, Sirotenko, Smith, Starkov, Stoecker, Sulkimo, Takahata,
  Tanaka, Tcherniaev, Safai~Tehrani, Tropeano, Truscott, Uno, Urban, Urban,
  Verderi, Walkden, Wander, Weber, Wellisch, Wenaus, Williams, Wright, Yamada,
  Yoshida, and Zschiesche]{agostinelli_geant4simulation_2003}
S.~Agostinelli, J.~Allison, K.~Amako, J.~Apostolakis, H.~Araujo, P.~Arce,
  M.~Asai, D.~Axen, S.~Banerjee, G.~Barrand, F.~Behner, L.~Bellagamba,
  J.~Boudreau, L.~Broglia, A.~Brunengo, H.~Burkhardt, S.~Chauvie, J.~Chuma,
  R.~Chytracek, G.~Cooperman, G.~Cosmo, P.~Degtyarenko, A.~Dell'Acqua,
  G.~Depaola, D.~Dietrich, R.~Enami, A.~Feliciello, C.~Ferguson, H.~Fesefeldt,
  G.~Folger, F.~Foppiano, A.~Forti, S.~Garelli, S.~Giani, R.~Giannitrapani,
  D.~Gibin, J.J. Gómez~Cadenas, I.~González, G.~Gracia~Abril, G.~Greeniaus,
  W.~Greiner, V.~Grichine, A.~Grossheim, S.~Guatelli, P.~Gumplinger,
  R.~Hamatsu, K.~Hashimoto, H.~Hasui, A.~Heikkinen, A.~Howard, V.~Ivanchenko,
  A.~Johnson, F.W. Jones, J.~Kallenbach, N.~Kanaya, M.~Kawabata, Y.~Kawabata,
  M.~Kawaguti, S.~Kelner, P.~Kent, A.~Kimura, T.~Kodama, R.~Kokoulin,
  M.~Kossov, H.~Kurashige, E.~Lamanna, T.~Lampén, V.~Lara, V.~Lefebure,
  F.~Lei, M.~Liendl, W.~Lockman, F.~Longo, S.~Magni, M.~Maire, E.~Medernach,
  K.~Minamimoto, P.~Mora~de Freitas, Y.~Morita, K.~Murakami, M.~Nagamatu,
  R.~Nartallo, P.~Nieminen, T.~Nishimura, K.~Ohtsubo, M.~Okamura, S.~O'Neale,
  Y.~Oohata, K.~Paech, J.~Perl, A.~Pfeiffer, M.G. Pia, F.~Ranjard, A.~Rybin,
  S.~Sadilov, E.~Di~Salvo, G.~Santin, T.~Sasaki, N.~Savvas, Y.~Sawada,
  S.~Scherer, S.~Sei, V.~Sirotenko, D.~Smith, N.~Starkov, H.~Stoecker,
  J.~Sulkimo, M.~Takahata, S.~Tanaka, E.~Tcherniaev, E.~Safai~Tehrani,
  M.~Tropeano, P.~Truscott, H.~Uno, L.~Urban, P.~Urban, M.~Verderi, A.~Walkden,
  W.~Wander, H.~Weber, J.P. Wellisch, T.~Wenaus, D.C. Williams, D.~Wright,
  T.~Yamada, H.~Yoshida, and D.~Zschiesche.
\newblock Geant4—a simulation toolkit.
\newblock \emph{Nuclear Instruments and Methods in Physics Research Section A:
  Accelerators, Spectrometers, Detectors and Associated Equipment},
  506\penalty0 (3):\penalty0 250--303, July 2003.
\newblock ISSN 01689002.
\newblock \doi{10.1016/S0168-9002(03)01368-8}.
\newblock URL
  \url{https://linkinghub.elsevier.com/retrieve/pii/S0168900203013688}.

\bibitem[Korcyl et~al.(2018)Korcyl, Białas, Curceanu, Czerwiński, Dulski,
  Flak, Gajos, Głowacz, Gorgol, Hiesmayr, Jasińska, Kacprzak, Kajetanowicz,
  Kisielewska, Kowalski, Kozik, Krawczyk, Krzemień, Kubicz, Mohammed,
  Niedźwiecki, Pawlik-Niedźwiecka, Pałka, Raczyński, Rajda, Rudy, Salabura,
  Sharma, Sharma, Shopa, Skurzok, Silarski, Strzempek, Wieczorek, Wiślicki,
  Zaleski, Zgardzińska, Zieliński, and Moskal]{korcyl}
G.~Korcyl, P.~Białas, C.~Curceanu, E.~Czerwiński, K.~Dulski, B.~Flak,
  A.~Gajos, B.~Głowacz, M.~Gorgol, B.~C. Hiesmayr, B.~Jasińska, K.~Kacprzak,
  M.~Kajetanowicz, D.~Kisielewska, P.~Kowalski, T.~Kozik, N.~Krawczyk,
  W.~Krzemień, E.~Kubicz, M.~Mohammed, Sz. Niedźwiecki,
  M.~Pawlik-Niedźwiecka, M.~Pałka, L.~Raczyński, P.~Rajda, Z.~Rudy,
  P.~Salabura, N.~G. Sharma, S.~Sharma, R.~Y. Shopa, M.~Skurzok, M.~Silarski,
  P.~Strzempek, A.~Wieczorek, W.~Wiślicki, R.~Zaleski, B.~Zgardzińska,
  M.~Zieliński, and P.~Moskal.
\newblock Evaluation of single-chip, real-time tomographic data processing on
  fpga soc devices.
\newblock \emph{IEEE Transactions on Medical Imaging}, 37\penalty0
  (11):\penalty0 2526--2535, 2018.
\newblock \doi{10.1109/TMI.2018.2837741}.

\bibitem[Gorgol et~al.(2020)Gorgol, Jasińska, Kosior, Stępień, and
  Moskal]{gorgol_construction_2020}
M.~Gorgol, B.~Jasińska, M.~Kosior, E.~Stępień, and P.~Moskal.
\newblock Construction of the {Vacuum} {Chambers} for {J}-{PET} {Experiments}
  with {Positron} {Annihilation}.
\newblock \emph{Acta Phys. Pol. B}, 51\penalty0 (1):\penalty0 293, 2020.
\newblock ISSN 0587-4254, 1509-5770.
\newblock \doi{10.5506/APhysPolB.51.293}.
\newblock URL
  \url{http://www.actaphys.uj.edu.pl/findarticle?series=Reg&vol=51&page=293}.

\bibitem[Krzemien et~al.(2020)Krzemien, Gajos, Kacprzak, Rakoczy, and
  Korcyl]{jpetFW}
W.~Krzemien, A.~Gajos, K.~Kacprzak, K.~Rakoczy, and G.~Korcyl.
\newblock J-{PET} {Framework}: {Software} platform for {PET} tomography data
  reconstruction and analysis.
\newblock \emph{SoftwareX}, 11:\penalty0 100487, January 2020.
\newblock ISSN 23527110.
\newblock \doi{10.1016/j.softx.2020.100487}.
\newblock URL
  \url{https://linkinghub.elsevier.com/retrieve/pii/S2352711020300509}.

\bibitem[Sharma et~al.(2020)Sharma, Chhokar, Curceanu, Czerwiński, Dadgar,
  Dulski, Gajewski, Gajos, Gorgol, Gupta-Sharma, Del~Grande, Hiesmayr,
  Jasińska, Kacprzak, Kapłon, Karimi, Kisielewska, Klimaszewski, Korcyl,
  Kowalski, Kozik, Krawczyk, Krzemień, Kubicz, Mohammed, Niedzwiecki, Pałka,
  Pawlik-Niedźwiecka, Raczyński, Raj, Ruciński, Shivani, Shopa, Silarski,
  Skurzok, Stępień, Wiślicki, Zgardzińska, and
  Moskal]{sharma_estimating_2020}
S.~Sharma, J.~Chhokar, C.~Curceanu, E.~Czerwiński, M.~Dadgar, K.~Dulski,
  J.~Gajewski, A.~Gajos, M.~Gorgol, N.~Gupta-Sharma, R.~Del~Grande, B.C.
  Hiesmayr, B.~Jasińska, K.~Kacprzak, {\L{}}.~Kapłon, H.~Karimi,
  D.~Kisielewska, K.~Klimaszewski, G.~Korcyl, P.~Kowalski, T.~Kozik,
  N.~Krawczyk, W.~Krzemień, E.~Kubicz, M.~Mohammed, Sz. Niedzwiecki,
  M.~Pałka, M.~Pawlik-Niedźwiecka, L.~Raczyński, J.~Raj, A.~Ruciński,
  S.~Shivani, R.Y. Shopa, M.~Silarski, M.~Skurzok, E.{\L{}}. Stępień,
  W.~Wiślicki, B.~Zgardzińska, and P.~Moskal.
\newblock Estimating relationship between the time over threshold and energy
  loss by photons in plastic scintillators used in the {J}-{PET} scanner.
\newblock \emph{EJNMMI Phys}, 7\penalty0 (1):\penalty0 39, December 2020.
\newblock ISSN 2197-7364.
\newblock \doi{10.1186/s40658-020-00306-x}.
\newblock URL
  \url{https://ejnmmiphys.springeropen.com/articles/10.1186/s40658-020-00306-x}.

\bibitem[Hiesmayr et~al.(2024)Hiesmayr, Krzemie{\'n}, and
  Ba{\l}a]{hiesmayrQuantumErrorChannels2024}
B.~C. Hiesmayr, W.~Krzemie{\'n}, and M.~Ba{\l}a.
\newblock Quantum error channels in high energetic photonic systems.
\newblock \emph{Sci Rep}, 14\penalty0 (1):\penalty0 9672, April 2024.
\newblock ISSN 2045-2322.
\newblock \doi{10.1038/s41598-024-60472-1}.

\bibitem[Caradonna(2024)]{caradonnaKinematicAnalysisMultiple2024}
Peter Caradonna.
\newblock Kinematic analysis of multiple compton scattering in
  quantum-entangled two-photon systems.
\newblock \emph{Annals of Physics}, 470:\penalty0 169779, November 2024.
\newblock ISSN 00034916.
\newblock \doi{10.1016/j.aop.2024.169779}.

\bibitem[Caradonna et~al.(2024)Caradonna, D'Amico, Jenkins, and
  Watts]{caradonnaStokesparameterRepresentationCompton2024}
P.~Caradonna, I.~D'Amico, D.~G. Jenkins, and D.~P. Watts.
\newblock Stokes-parameter representation for compton scattering of entangled
  and classically correlated two-photon systems.
\newblock \emph{Phys. Rev. A}, 109\penalty0 (3):\penalty0 033719, March 2024.
\newblock ISSN 2469-9926, 2469-9934.
\newblock \doi{10.1103/PhysRevA.109.033719}.

\bibitem[Tkachev et~al.(2025)Tkachev, Musin, Abdurashitov, Baranov, Guber,
  Ivashkin, and Strizhak]{tkachevMeasuringEvolutionEntanglement2025}
Igor Tkachev, Sultan Musin, Dzhonrid Abdurashitov, Alexander Baranov, Fedor
  Guber, Alexander Ivashkin, and Alexander Strizhak.
\newblock Measuring the evolution of entanglement in compton scattering.
\newblock \emph{Sci Rep}, 15\penalty0 (1):\penalty0 6064, February 2025.
\newblock ISSN 2045-2322.
\newblock \doi{10.1038/s41598-025-87095-4}.

\bibitem[Hiesmayr and Moskal(2017)]{hiesmayr_genuine_2017}
Beatrix~C. Hiesmayr and Pawel Moskal.
\newblock Genuine {Multipartite} {Entanglement} in the 3-{Photon} {Decay} of
  {Positronium}.
\newblock \emph{Sci Rep}, 7\penalty0 (1):\penalty0 15349, November 2017.
\newblock ISSN 2045-2322.
\newblock \doi{10.1038/s41598-017-15356-y}.
\newblock URL \url{https://www.nature.com/articles/s41598-017-15356-y}.

\end{thebibliography}

\appendix
\section{Additional selection criteria for annihilation hits}
\label{appendix:annihilation_selection}
The candidate for the {\it annihilation hit} is a reconstructed point of interaction with the detector associated with the primary interaction of the annihilation photon based on the TOT selection. Since we consider the double-scattered photons, we expect two annihilation hits in the four-hit event. To reduce the background contribution, we apply additional constraints to the annihilation pair candidates, taking advantage of the fact that the emission area of the back-to-back 511 keV pairs is reduced to about 1 cm$^{3}$  around the positronium source placed in the geometrical centre of the detector.
We apply the selection conditions based on the registered time difference and hit positions that limit the reconstructed annihilation point to the vicinity of the geometrical centre.
Namely, the registration time difference $\Delta t$ must be smaller than $1$~ns. Also,  the reconstructed annihilation vertex $(x_a,y_a,z_a)$ is restricted by the following conditions: $x_a^2 +y_a^2 < 10~\text{cm}^2$ and $|z_a|<4~\text{cm}$ (see \refFig{fig:emission-point-xy-histograms} and \refFig{fig:emission-point-z-histograms}, respectively). The separation of $x_a, y_a$ and $z_a$ selection conditions is due to different experimental resolutions.
\addTwoFigsH
{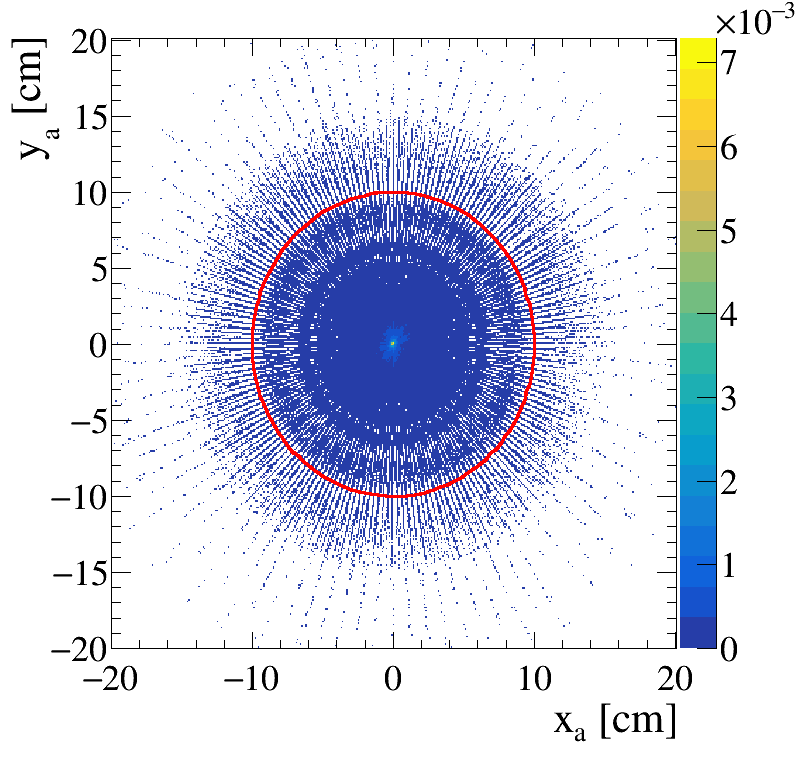}{Experimental data}{fig:emission-point-xy-histograms:experiment}
{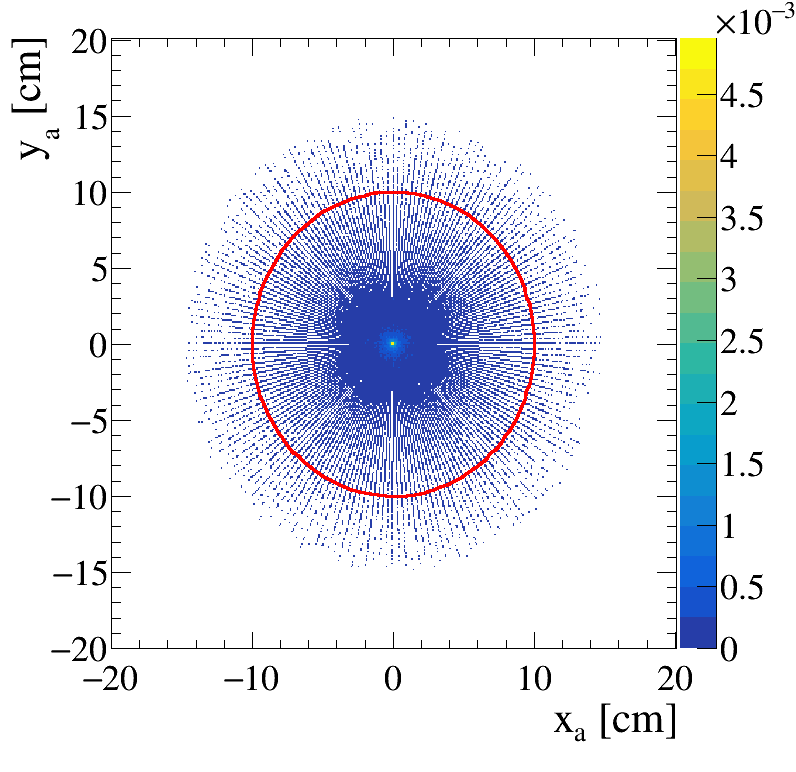}{MC simulations}{fig:emission-point-xy-histograms:simulation}
{Reconstructed annihilation points at XY plane. The red circle shows the selection region $x_a^2 +y_a^2 < 10~\text{cm}^2$. The visible pattern is due to the geometrical acceptance of the detector. The spectra are normalized to unity.}
{fig:emission-point-xy-histograms}{0.5}
\addTwoFigsH
{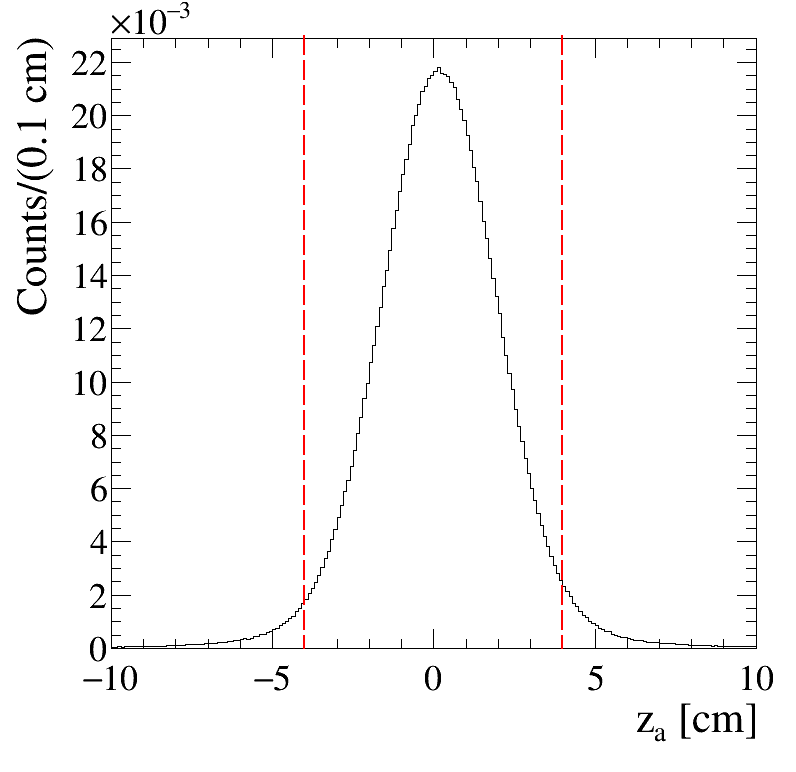}{Experimental data}{fig:emission-point-z-histograms:experiment}
{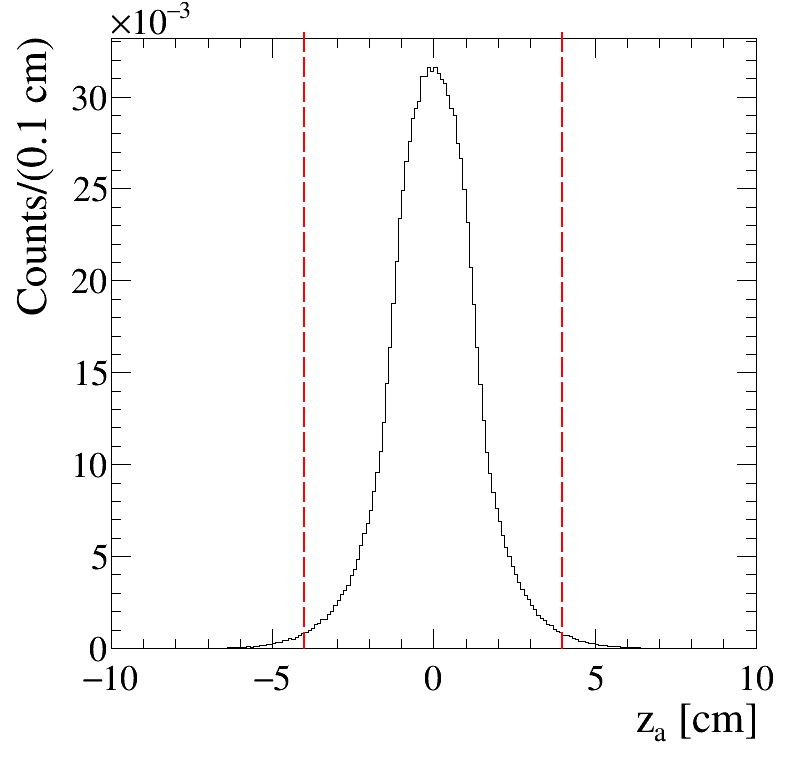}{MC simulations}{fig:emission-point-z-histograms:simulation}
{Reconstructed annihilation point in Z dimension. Red dashed lines show the selection region $|z_a|<4~\text{cm}$. The spectra are normalized to unity.}
{fig:emission-point-z-histograms}{0.5}
The applied selection criteria retain 95\% of the signal events.

\section{Calculation of the weighted mean of the visibility squared over the selection region}
\label{appendix:weighted_mean}
Let us consider a set of polarized two-photon pairs (e.g. in the initial state $\ket{\Psi^{+}}$) measured by a perfect detector. For fixed scattering angles $(\tTheta_1, \tTheta_2)$, the conditional probability density function describing the measurement of the configuration $\DeltaPhi$ will be given by:
\begin{equation}
\label{eq:hDeltaPhi}
h(\DeltaPhi|\tTheta_1, \tTheta_2) = \frac{1}{\pi}(1 + \Vis(\tTheta_1)\Vis(\tTheta_2)\,\cos(2\DeltaPhi))
\end{equation}
Now, we consider a range of scattering angles confined to a region (see e.g. \refFig{fig:experimetaldata:tvt}) with the density of the measured events given by $\rho(\tTheta_1, \tTheta_2)$. The conditional probability density function averaged over the selected region will be equal to:

\OneLineEq{eq:hist-mean-value-region}{\Mean{h(\DeltaPhi)}_{region} = \int\limits_{region}\rho(\tTheta_1, \tTheta_2)h(\DeltaPhi|\tTheta_1, \tTheta_2) d\tTheta_1 d\tTheta_2,}
which leads to: 

\begin{equation}
\Mean{h(\DeltaPhi)}_{region} = \frac{1}{\pi}(1+\Mean{\Vis^{2}}_{region}\cos(2\DeltaPhi))
\end{equation}
The last equation can be interpreted as follows: if we form the $\DeltaPhi$ distribution from all measured events in the selection area, and fit it with the function $f(\DeltaPhi) = a(1+b \cos(2\DeltaPhi))$,  then the $b$ parameter of the fit would correspond to the averaged visibility squared $\Mean{\Vis^{2}}_{region}$ calculated over the selection region.
Note that in this reasoning only $\DeltaPhi$ is treated as a random variable, while $\tTheta_1, \tTheta_2$ are treated as parameters, and the density $\rho(\tTheta_1, \tTheta_2)$ as fixed corresponding to the density of the measured events.
The $\Mean{\Vis^{2}}_{circle}$ factor is given by the equation:
\begin{equation}
\label{eq:vis2integral}
\Mean{\Vis^{2}}_{circle}  = \int\limits_{C(\tTheta_a,\tTheta_b,R_\tTheta)} \rho(\tTheta_1, \tTheta_2) \Vis(\tTheta_1) \Vis(\tTheta_2) d\tTheta_1 d\tTheta_2
\end{equation}
where $C(\tTheta_a,\tTheta_b,R_\tTheta)$ is a circle defined by the inequality:
\begin{equation}
(\tTheta_1-\tTheta_a)^2 +(\tTheta_2-\tTheta_b)^2 \leq R_\tTheta^2
\end{equation}

To determine  $\Mean{\Vis^{2}}_{circle}$ from the data, one can use the discretized version of \refEq{eq:vis2integral}, and calculate the weighted mean over the selection region, where weighs are formed from the $n_{ij}$ corresponding to the registered number of events  in the $i,j$-th bin:
\ManyLinesEq{eq:approx-region-mean-vis2}{\Mean{\Vis^{2}}_{circle}  \approx \sum_{i=1}^{N_i}\sum_{j=1}^{N_j} \delta(i,j,R_\tTheta) \frac{n_{ij}}{N_i N_j}\Vis(\tTheta_1(i))\Vis(\tTheta_2(j)) \Delta \tTheta_1\Delta \tTheta_2}

\begin{equation}
\delta(i,j,R_\tTheta) = \begin{cases}
1,~(\tTheta_1(i),\tTheta_2(j))\in C(\tTheta_a,\tTheta_b,R_\tTheta),\\
0,\text{otherwise}
\end{cases}
\end{equation}

Equivalently, $\Mean{\Vis^{2}}_{circle}$ can be also determined using the {\it unbinned} approach:
\begin{equation}
\langle\Vis^{2}\rangle_{circle}  \approx \frac{1}{N_{\text{events}}}\sum_{\text{event}=1}^{N_{\text{events}}} \Vis(\tTheta_1(\text{event}))\Vis(\tTheta_2(\text{event}))
\end{equation}

In summary, the $\Mean{\Vis^{2}}_{circle}$ factor can be extracted from the experimental distribution and plays the role of an upper bound of the squared visibility value that can potentially be determined by fitting the $\DeltaPhi$ distribution for the given selection region. In practice, the visibility can be lower due to the experimental dilution effects, the unpolarized background, or dependence on the initial polarization state.

\section{Estimation of the fit bias influence}
\label{appendix:fit_systematics}

The fit to the experimental data is slightly biased as seen in the pull distribution in ~\refFig{fig:correction-r1-fits}, bottom panel. 
Namely, near the $\DeltaPhi=0^{\deg}$ the fitted function passes on average below the data points, while for the $\DeltaPhi=180^{\deg}$ the effect is reversed. The presence of the bias might be the reflection of some of the experimental effects which are not reproduced by the MC simulations and therefore cannot be compensated by the correction procedure.     
To quantify the observed bias, we fit the $\DeltaPhi$ distribution by a function $g(\DeltaPhi) = a(1 + b\,\cos(2(\DeltaPhi +c))$, where the introduced $c$ parameter accounts for the possible phase shift from the nominal $90 ^{\deg}$ center. The results of the fits to both experimental and MC distributions are shown in \refFig{fig:phase-shift}.
\addTwoFigsH
{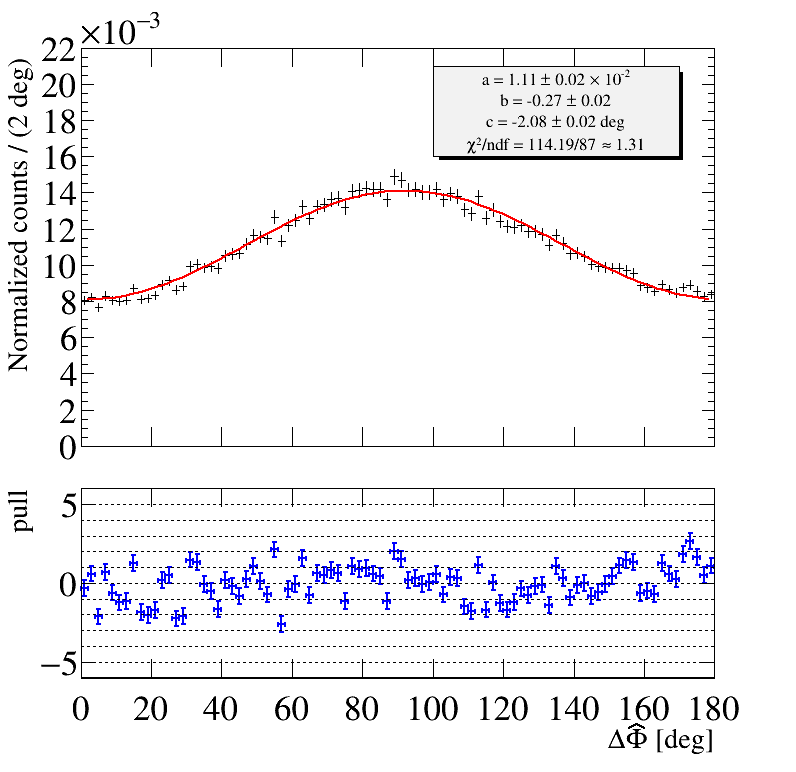}{Experimental data}{fig:phase-shift:exp}
{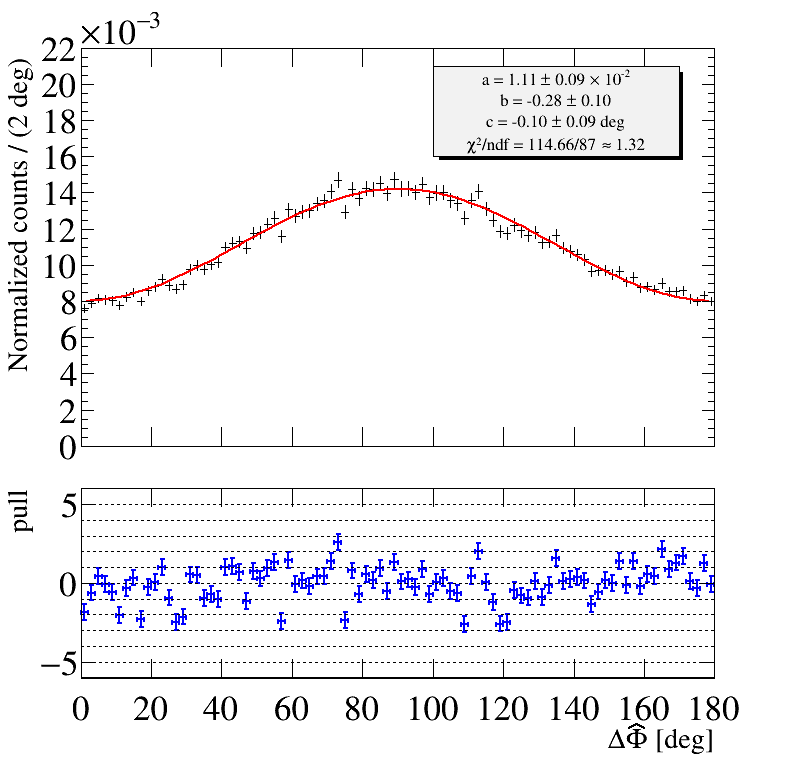}{MC simulations}{fig:phase-shift:sim}
{The comparison of experimental and MC $\DeltaPhi$ distributions. The spectra are normalized to unity. The distributions are fitted with the function (red line)   $g(\DeltaPhi) = a(1+b\cos(2(\DeltaPhi +c)))$ with the reduced chi-2 equal to $\chi^{2}_{exp} = 114.19/87 \approx 1.31$ and $\chi^{2}_{MC} = 114.66/87 \approx 1.32$ for data and MC respectively. The MC distribution is in good agreement with the data distribution. The fitted values of $a$ and $b$ parameters are equivalent within the statistical errors. The fit to the experimental data reveals the $2.08^{\deg}$ shift from the nominal $90^{\deg}$ distribution center. }
{fig:phase-shift}{0.5}
Indeed, the shift $c = -2.08 \pm 0.02 ^{\deg}$ is statistically important for the experimental distribution. However,  the presence of the bias does not change the values of the visibility squared parameter, which remains the same both for data and MC fits.

\end{document}